\documentclass[twocolumn,american,prb,showpacs]{revtex4}
\usepackage{pslatex}
\usepackage[T1]{fontenc}
\usepackage[latin1]{inputenc}
\usepackage{amsmath}
\usepackage{graphicx}
\usepackage{amssymb}

\makeatletter
\usepackage{nicefrac}

\usepackage{babel}
\makeatother
\begin{document}

\title{Aperiodic quantum \emph{XXZ} chains: Renormalization-group results}

\author{Andr\'{e} P. Vieira}

\affiliation{Instituto de F\'{\i}sica da Universidade de S\~{a}o Paulo, Caixa
Postal 66318, 05315-970, S\~{a}o Paulo, Brazil}

\email{apvieira@if.usp.br}

\date{\today{}}


\begin{abstract}
We report a comprehensive investigation of the low-energy properties
of antiferromagnetic quantum \emph{XXZ} spin chains with aperiodic
couplings. We use an adaptation of the Ma-Dasgupta-Hu renormalization-group
method to obtain analytical and numerical results for the low-temperature
thermodynamics and the ground-state correlations of chains with couplings
following several two-letter aperiodic sequences, including the quasiperiodic
Fibonacci and other precious-mean sequences, as well as sequences
inducing strong geometrical fluctuations. For a given aperiodic sequence,
we argue that in the easy-plane anisotropy regime, intermediate between
the \emph{XX} and Heisenberg limits, the general scaling form of the
thermodynamic properties is essentially given by the exactly-known
\emph{XX} behavior, providing a classification of the effects of aperiodicity
on \emph{XXZ} chains. We also discuss the nature of the ground-state
structures, and their comparison with the random-singlet phase, characteristic
of random-bond chains.
\end{abstract}
\maketitle

\section{Introduction}

At low temperatures, the interplay between lack of translational invariance
and quantum fluctuations in low-dimensional strongly-correlated electron
systems may induce novel phases with peculiar behavior. In particular,
randomness in quantum spin chains may lead to Griffiths phases,\cite{fisher92,fisher95}
random quantum paramagnetism,\cite{furusaki94,furusaki95,nguyen96}
large-spin formation,\cite{westerberg95,westerberg97} and random-singlet
phases.\cite{fisher94,refael02} On the other hand, studies on the
influence of deterministic but aperiodic elements on similar systems
(see e.g. Refs. \onlinecite{kohmoto83, luck86, luck93a, vidal99, vidal01, hida99, hida01, hermisson00, arlego01}),
inspired by the experimental discovery of quasicrystals,\cite{shechtman84}
have revealed strong effects on dynamical and thermodynamic properties,
although much less is known concerning the precise nature of the underlying
ground-state phases.%
\footnote{ For recent results on two-dimensional antiferromagnetic quasycristals
see S. Wessel, A. Jagannathan, and S. Haas, Phys. Rev. Lett. \textbf{90},
177205 (2003), and also A. Jagannathan, Phys. Rev. Lett. \textbf{92},
047202 (2004); preprint: cond-mat/0409711.%
}

Prototypical models for those studies are spin-$\nicefrac{1}{2}$
antiferromagnetic \emph{XXZ} chains described by the Hamiltonian\begin{equation}
H=\sum_{i}J_{i}\left(S_{i}^{x}S_{i+1}^{x}+S_{i}^{y}S_{i+1}^{y}+\Delta S_{i}^{z}S_{i+1}^{z}\right),\end{equation}
 where $J_{i}>0$ and the $S_{i}$ are spin operators. In the uniform
case ($J_{i}\equiv J$), the ground state for chains with $-1<\Delta\leq1$
is critical,\cite{baxter72} exhibiting power-law decay of the pair
correlations as a function of the distance between spins,\cite{luther75}
as well as gapless elementary excitations. Such critical phase is
unstable towards dimerization, i.e. the introduction of alternating
couplings $J_{\mathrm{{odd}}}$ and $J_{\mathrm{{even}}}$, in the
presence of which a gap opens between the (now localized) ground state
and the first excitated states.\cite{cross79,bonner82,barnes99} This
instability hints at the profound effects produced by fully breaking
the translational symmetry of the system.

Random-bond versions of these chains have been much studied by a real-space
renormalization-group (RG) method introduced\cite{ma79,dasgupta80}
by Ma, Dasgupta and Hu (MDH) for the Heisenberg chain ($\Delta=1$)
and more recently extended by Fisher,\cite{fisher92,fisher94,fisher95}
who gave evidence that the method becomes asymptotically exact at
low energies. In the last few years, the method has been applied and
adapted to a variety of random systems (see e.g. Refs. \onlinecite{westerberg95, westerberg97, hyman96, doussal99, saguia02, yusuf02, hoyos04a, hooyberghs03, hoyos04}).
The basic idea is to decimate the spin pairs coupled by the strongest
bonds (those with the largest gaps between the local ground state
and the first excited multiplet), forming singlets and inducing weak
effective couplings between neighboring spins, thereby reducing the
energy scale. For \emph{XXZ} chains in the regime $-\nicefrac{1}{2}<\Delta\leq1$,
the method predicts the ground state to be a random-singlet phase,
consisting of arbitrarily distant spins forming rare, strongly-correlated
singlet pairs.\cite{fisher94} 

Another way of breaking the translational symmetry is suggested by
analogies with quasicrystals. These are structures which exhibit symmetries
forbidden by traditional crystalography, and which correspond to projections
of higher-dimensional Bravais lattices onto low-dimensional subspaces.\cite{janner77}
A one-dimensional example is provided by the Fibonacci quasiperiodic
chain, obtained from a cut-and-project operation on a square lattice.\cite{elser85}
In this chain, spins are separated by two possible distances, $a$
and $b$, whose sequence, starting from the left end of the chain,
is $abaab\ldots$ This sequence can be generated by repeatedly applying
a substitution (or inflation) rule $a\rightarrow ab$, $b\rightarrow a$,
starting from a single distance $a$. Associating with each $a$ a
coupling $J_{a}$ and with each $b$ a coupling $J_{b}$ we obtain
a spin chain with couplings following a Fibonacci sequence. More generally,
we can postulate a two-letter substitution rule, build the corresponding
letter sequence, and associate couplings with letters to obtain spin
chains whose couplings follow aperiodic but deterministic sequences.%
\footnote{There are however cases where a substitution rule generates a periodic
sequence, as in $a\rightarrow ab$, $b\rightarrow ab$. It is possible
to establish conditions under which a substitution rule generates
an aperiodic sequence; see e.g. S. T. R. Pinho and T. C. P. Lobão,
Braz. J. Phys. \textbf{30}, 772 (2000).%
} Quasiperiodic sequences are characterized by a Fourier spectrum consisting
of Bragg peaks, but more complex spectra (such as singular continuous)
can be generated by substitution rules.\cite{godreche92} In this
work we apply the term `aperiodic' when referring to nonperiodic,
self-similar sequences, also encompassing those which are strictly
quasiperiodic in the above sense.

In \emph{XX} spin chains ($\Delta=0$), the low-temperature thermodynamic
behavior can be qualitatively determined for virtually any aperiodic
sequence by an exact RG method.\cite{hermisson00} The effects of
aperiodicity depend on topological properties of the sequence. If
the fraction of letters $a$ (or $b$) at odd positions is different
from that at even positions (i.e. if there is average dimerization),
then a finite gap opens between the global ground state and the first
excited states, and the chain becomes noncritical. Otherwise, the
scaling of the lowest gaps can be classified according to the wandering
exponent $\omega$ measuring the geometric fluctuations $g$ related
to nonoverlapping pairs of letters,\cite{hermisson00} which vary
with the system size $N$ as $g\sim N^{\omega}$. If $\omega<0$,
aperiodicity has no effect on the long-distance, low-temperature properties,
and the system behaves as in the uniform case, with a finite susceptibility
at $T=0$. If $\omega=0$, as in the Fibonacci sequence, aperiodicity
is marginal and may lead to nonuniversal power-law scaling behavior
of thermodynamic properties. If $\omega>0$, aperiodicity is relevant
in the RG sense, affecting the $T=0$ critical behavior and leading
to exponential scaling of the lowest gaps $\Lambda$ at long distances
$r$, according to the form $\Lambda\sim\exp\left(-r^{\omega}\right)$.
In particular, for sequences with $\omega=\nicefrac{1}{2}$, geometric
fluctuations mimic those induced by randomness, and the scaling behavior
is similar to the one characterizing the random-singlet phase.\cite{fisher94} 

In contrast, results for the effects of aperiodicity on low-energy
properties of \emph{XXZ} chains have been so far scarce, and restricted
to particular sequences. Vidal, Mouhanna and Giamarchi\cite{vidal99,vidal01}
studied the related problem of an interacting spinless fermion chain
with Fibonacci or precious-mean potential by using bosonization techniques,
which are valid in the weak-modulation regime ($J_{a}\simeq J_{b}$).
At half filling, where the system corresponds to an \emph{XXZ} chain
in zero external field, their calculations predict that aperiodicity
will drive the system away from the usual Luttinger-liquid behavior
for $0\leq\Delta\leq1$. A similar conclusion is drawn from studies
on a Hubbard chain with hoppings following a Fibonacci sequence.\cite{hida01}
Density-matrix renormalization-group (DMRG) results on \emph{XXZ}
chains with precious-mean couplings,\cite{hida99} and recent real-space
RG calculations on the Fibonacci \emph{XXZ} chain\cite{hida04a,hida04b}
(also based on the MDH scheme), likewise predict that low-temperature
properties are different than in the uniform chains. The zero-temperature
magnetization curve of Fibonacci \emph{XXZ} chains has also been investigated,\cite{arlego01}
with emphasis on determining the plateau structure.

Our aim in this paper is to investigate the effects of \emph{arbitrary}
aperiodic coupling distributions on the low-temperature properties
of \emph{XXZ} chains, reinforcing and extending our previous results.\cite{vieira04}
From an adaptation of the Ma-Dasgupta-Hu RG scheme, we obtain information
about low-temperature thermodynamics and ground-state correlation
functions for several aperiodic sequences. Our results, which are
presumably exact in the strong-modulation limit ($J_{a}\ll J_{b}$
or $J_{a}\gg J_{b}$), point to the following conclusions:

\begin{itemize}
\item The exact classification found in the \emph{XX} limit can arguably
be extended to \emph{XXZ} chains in the anisotropy regime $0<\Delta\leq1$.
We predict that dimerized aperiodicity opens a gap to the lowest excitations,
and that otherwise the effects of aperiodicity on the low-temperature
thermodynamics are gauged by the same exponent $\omega$, irrespective
of anisotropy. In particular, sequences which are strictly marginal
in the \emph{XX} limit continue to be so for anisotropies $0<\Delta<1$,
but may be marginally relevant in the Heisenberg limit.
\item On the other hand, $\omega$ is found not to define the behavior of
correlation functions, although ground-state structures in the presence
of marginal or relevant couplings also reflect self-similar properties
of the sequences. Dominant correlations correspond to well-defined
distances, related to the rescaling factor of the sequences (contrary
to the random-singlet phase, where no such characteristic distances
exist), and two types of behavior are possible: either the chains
can be decomposed into a hierarchy of singlets, forming a kind of
`aperiodic-singlet phase', or into a hierarchy of effective spins,
in which case low-energy excitations involve an exponentially large
number of spins. This is in sharp contrast both to the gapless spin-wave
excitations of the uniform chains and to the gapped triplet-wave excitations
of the dimerized chains. 
\item Based on second-order calculations, the long-distance decay exponents
of average ground-state correlation functions are seen to vary with
the coupling ratio in the presence of strictly marginal aperiodicity.
Otherwise, strong universality (i.e. independence of the exponents
on both coupling ratio and anisotropy) is obtained for the whole line
$0\leq\Delta<1$, although different decay exponents may emerge in
the Heisenberg limit. Also, the scaling form of typical (rather than
average) correlations follows essentially the same scaling form as
the energy gaps, similarly to what happens for random-bond chains.
\end{itemize}

In order to make the paper self-contained, we begin by reviewing some
known results. So, in Sec. \ref{sec:randombond} we present the basics
of the Ma-Dasgupta-Hu scheme, as applied to random-bond \emph{XXZ}
chains, and summarize the properties of the underlying random-singlet
phase. Also, in Sec. \ref{sec:aperxx} we provide a short discussion
on aperiodic sequences, as well as a sketch of the exact RG results
for \emph{XX} chains with aperiodic couplings. Our adaptation of the
Ma-Dasgupta-Hu method to aperiodic \emph{XXZ} chains is described
in Sec. \ref{sec:mdhxxz}, and results for marginal and relevant aperiodicity
are presented in Secs. \ref{sec:marg} and \ref{sec:rel}. The final
section is devoted to a discussion and conclusions. There are also
two appendices, in which some important technical points are detailed.

\section{Random-bond spin chains and the Ma-Dasgupta-Hu method\label{sec:randombond}}

Consider an antiferromagnetic quantum spin-$\nicefrac{1}{2}$ chain
described by the Hamiltonian\begin{equation}
H=\sum_{i}J_{i}\left(S_{i}^{x}S_{i+1}^{x}+S_{i}^{y}S_{i+1}^{y}+\Delta_{i}S_{i}^{z}S_{i+1}^{z}\right),\label{eq:xxzhamilt}\end{equation}
where $J_{i}>0$ and all anisotropies are such that $0\leq\Delta_{i}\leq1$.
Let us assume that the couplings $J_{i}$ are randomly distributed
according to a broad probability distribution $\wp(J_{i})$ having
an upper cutoff $J_{\textrm{max}}$. Under such conditions, in a finite
but large chain, there is a strongest bond $J_{0}\simeq J_{\textrm{max}}$
connecting, say, spins $S_{1}$ and $S_{2}$, which on their turn
are coupled to their other nearest neighbors $S_{l}$ and $S_{r}$
by weaker bonds $J_{l}$ and $J_{r}$. The local Hamiltonian connecting
$S_{1}$ and $S_{2}$ is\[
H_{0}=J_{0}\left(S_{1}^{x}S_{2}^{x}+S_{1}^{y}S_{2}^{y}+\Delta_{0}S_{1}^{z}S_{2}^{z}\right),\]
whose ground state is a singlet, separated from the first excited
states by an energy gap $\Lambda_{0}=\tfrac{1}{2}(1+\Delta_{0})J_{0}$.
The idea behind the Ma-Dasgupta-Hu\cite{ma79,dasgupta80} method is
that, at temperatures below $\Lambda_{0}$, $S_{1}$ and $S_{2}$
can be decimated out of the system, since they couple into a singlet,
giving a negligible contribution to thermodynamic properties. Nevertheless,
their virtual excitations induce a weak effective coupling between
$S_{l}$ and $S_{r}$, described by the Hamiltonian\[
H^{\prime}=J^{\prime}\left(S_{l}^{x}S_{r}^{x}+S_{l}^{y}S_{r}^{y}+\Delta^{\prime}S_{l}^{z}S_{r}^{z}\right).\]
The parameters $J^{\prime}$ and $\Delta^{\prime}$ can be obtained
by second-order perturbation theory (see Appendix \ref{sec:multi}),
and are given by\begin{equation}
J^{\prime}=\frac{1}{1+\Delta_{0}}\cdot\frac{J_{l}J_{r}}{J_{0}}\quad\textrm{and}\quad\Delta^{\prime}=\frac{1+\Delta_{0}}{2}\Delta_{l}\Delta_{r}.\label{eq:mdh2}\end{equation}
Notice that $J^{\prime}$ is smaller than either $J_{l}$, $J_{r}$
or $J_{0}$; likewise, unless all $\Delta_{i}=1$, $\Delta^{\prime}$
is smaller than either $\Delta_{l}$ or $\Delta_{r}$. Thus, after
eliminating $S_{1}$ and $S_{2}$, the overall energy scale is reduced.
The previous steps can be repeated with the next largest bond, which
most probably is not $J^{\prime}$. 

Starting from Eqs. (\ref{eq:mdh2}), Fisher\cite{fisher94} was able
to write and solve recursion relations for the probability distribution
of the effective couplings. The fixed-point distribution is presumably
independent of the initial couplings,\cite{laflorencie03} and diverges
as a power law for $J^{\prime}\to0^{+}$, indicating that the perturbative
approach leading to Eqs. (\ref{eq:mdh2}) becomes essentially exact
for asymptotically low energies.

The ground state is a `random-singlet' phase, consisting of arbitrarily
distant spins forming rare, strongly correlated singlet pairs.\cite{doty92,fisher94}
Exciting a singlet whose spins are separated by a distance $r$ costs
an energy of order $\Lambda$, with a dynamic scaling form\[
\Lambda\sim e^{-\mu\sqrt{r/r_{0}}},\]
where $\mu$ and $r_{0}$ are constants. At low temperatures, the
zero-field susceptibility diverges as\[
\chi\sim\frac{1}{T\ln^{2}T}.\]
Average ground-state correlations are dominated by the rare singlet
pairs, and decay as a power law,\[
C(r)\equiv\left|\overline{\left\langle \mathbf{S}_{i}\cdot\mathbf{S}_{i+r}\right\rangle }\right|\sim\frac{1}{r^{2}},\]
where the bar denotes average over the whole chain, while typical
correlations are short-ranged, following\[
C_{\mathrm{typ}}(r)\sim e^{-\mu_{\mathrm{typ}}\sqrt{r/r_{0}}}.\]
These asymptotic results are independent of the anisotropies $\Delta_{i}$,
as long as $0\leq\Delta_{i}\leq1$ for all $i$, with the same distribution
on even and odd bonds.

\section{Aperiodic sequences and \emph{XX} chains\label{sec:aperxx}}

Following closely the analysis of Hermisson,\cite{hermisson00} in
this Section we consider antiferromagnetic quantum \emph{XX} chains,
described by the Hamiltonian\begin{equation}
H=\sum_{i}J_{i}\left(S_{i}^{x}S_{i+1}^{x}+S_{i}^{y}S_{i+1}^{y}\right),\label{eq:xxhamilt}\end{equation}
where now the strengths of the site-dependent couplings $J_{i}$ can
be either $J_{a}$ or $J_{b}$, and are distributed according to deterministic
but aperiodic binary sequences, obtained by substitution (or inflation)
rules of the form\[
\sigma:\ \left\{ \begin{array}{c}
a\to w_{a}\\
b\to w_{b}\end{array}\right.,\]
where $w_{a}$ and $w_{b}$ are words (finite strings) composed of
letters $a$ and $b$. A well-know example is provided by the Fibonacci
sequence, whose substitution rule is\[
\sigma_{\textrm{fb}}:\left\{ \begin{array}{l}
a\to ab\\
b\to a\end{array}\right..\]
Starting from a single letter $a$, repeated application of $\sigma_{\textrm{fb}}$
yields strings with lengths given by the Fibonacci numbers $1$, $2$,
$3$, $5$, $8$,$\ldots$, ultimately producing a letter sequence
$abaababaaba\ldots$, for which no period can be identified.

Given an inflation rule $\sigma$, various statistical properties\cite{queffelec87}
of the associated sequence are enclosed in the substitution matrix\[
\mathbf{M}=\left(\begin{array}{cc}
\#_{a}(w_{a}) & \#_{a}(w_{b})\\
\#_{b}(w_{a}) & \#_{b}(w_{b})\end{array}\right),\]
where $\#_{a}(w_{b})$ denotes the number of letters $a$ in the word
$w_{b}$. The largest eigenvalue of $\mathbf{M}$, $\lambda_{+}$,
gives the asymptotic scaling factor of the string length (i.e. the
ratio between the lengths of the strings corresponding to successive
iterations of the rule $\sigma$); the entries of the corresponding
eigenvector are proportional to the frequencies $p_{a}$ and $p_{b}$
of letters $a$ and $b$ in the limit (infinite) sequence. 

The remaining eigenvalue, $\lambda_{-}$, is related to the geometric
fluctuations of the sequence, which are defined in the following way.
Let $N_{n}^{a}$ be the number of letters $a$ in the string obtained
after $n$ iterations of $\sigma$, and $N_{n}$ be the corresponding
total number of letters (the length of the string). Then, a measure
of the geometric fluctuations induced by the sequence is the difference
$g_{n}$ between $N_{n}^{a}$ and the number of letters $a$ expected
from the limit-sequence frequency $p_{a}$, and this behaves as \[
g_{n}=\left|N_{n}^{a}-p_{a}N_{n}\right|\sim\left|\lambda_{-}\right|^{n}.\]
Since $N_{n}\sim\lambda_{+}^{n}$, this last equation can be rewritten
as\[
g_{n}\sim N_{n}^{\omega_{l}},\]
by defining the `wandering exponent' \[
\omega_{l}=\frac{\ln\left|\lambda_{-}\right|}{\ln\lambda_{+}}.\]
If $\omega_{l}<0$, fluctuations become smaller as the string grows,
and the sequence looks more and more `periodic'. On the other hand,
if $\omega_{l}>0$, fluctuations increase without limit. The marginal
case $\omega_{l}=0$ is in general connected to logarithmic fluctuations.
It can be shown\cite{luck93c} that substitutions for which $\omega_{l}<0$
generate quasiperiodic (or limit-quasiperiodic) sequences.

The concept of geometric fluctuations is essential in establishing
the Harris-Luck criterion\cite{harris74,luck93b} for the relevance
of inhomogeneities to the critical behavior of magnetic systems. The
criterion states that, if fluctuations in the local parameters controlling
the criticality of the system vary with some characteristic length
$L$ as $g\sim L^{\omega}$, then there is a critical value of the
exponent $\omega$ above which the presence of inhomogeneities can
affect the critical behavior. This happens for\begin{equation}
\omega>1-\frac{1}{d\nu},\label{eq:harrisluck}\end{equation}
where $d$ is the number of dimensions along which inhomogeneities
are distributed, and $\nu$ is the correlation-length critical exponent
of the underlying uniform system. Randomly distributed inhomogeneities
lead to $\omega=\nicefrac{1}{2}$, and the general Harris criterion\cite{harris74,lubensky75,muzy02}
is recovered. 

The ground-state of the model in Eq. (\ref{eq:xxhamilt}) is critical
in the uniform limit ($J_{i}\equiv J$): there is no energy gap to
the low-lying excitations, and pair correlations decay as power laws,\cite{lieb61}\[
C^{\alpha\alpha}(r)=\left|\overline{\left\langle S_{i}^{\alpha}S_{i+r}^{\alpha}\right\rangle }\right|\sim r^{-\eta^{\alpha\alpha}},\]
 with $\eta^{xx}=\eta^{yy}=\nicefrac{1}{2}$ and $\eta^{zz}=2$. This
phase is unstable towards dimerization (i.e. the presence of couplings
$J_{\textrm{o}}$ and $J_{\textrm{e}}$ alternating between odd and
even bonds), in which case a gap opens in the low-energy spectrum,
and ground-state correlations become short-ranged. More generally,
the model exhibits a (zero-temperature) quantum phase transition between
two dimer phases for\cite{pfeuty79,hermisson00}\begin{equation}
\delta=\overline{\ln J_{2j-1}}-\overline{\ln J_{2j}}=0.\label{eq:critcond}\end{equation}
 If $J_{2j-1}\equiv J_{\textrm{o}}$ and $J_{2j}\equiv J_{\textrm{e}}$,
the phase transition occurs for $\delta=\ln(J_{\textrm{o}}/J_{\textrm{e}})=0$,
and belongs to the Onsager universality class, with $\nu=1$. 

When the couplings $J_{i}$ are chosen according to aperiodic sequences
for which the fractions of letters $a$ (or $b$) at even and odd
positions are different, Eq. (\ref{eq:critcond}) is not satisfied,
and the system is in a dimer phase. This suggests that the local parameters
defining the criticality of \emph{XX} chains are the shifts $\delta_{j}=\ln(J_{2j-1}/J_{2j})$.
In order to study the fluctuations of the $\delta_{j}$, which depend
on two consecutive couplings, we must usually consider the sequence
of nonoverlapping letter pairs associated with a given aperiodic sequence.
To build the inflation rule $\sigma^{(2)}$ for such pairs, it is
necessary to iterate the original rule $\sigma$ until the strings
obtained from a single $a$ and $b$ have lengths of the same parity.
As an illustration, let us take the Fibonacci sequence. Applying $\sigma_{\textrm{fb}}$
three times yields\[
\sigma_{\textrm{fb}}^{3}:\left\{ \begin{array}{l}
a\to abaab\\
b\to aba\end{array}\right..\]
Noting that the pair $bb$ does not occur in the sequence, we readily
obtain \[
\sigma_{\textrm{fb}}^{(2)}:\left\{ \begin{array}{l}
aa\to(ab)(aa)(ba)(ba)(ab)\\
ab\to(ab)(aa)(ba)(ba)\\
ba\to(ab)(aa)(ba)(ab)\end{array}\right..\]

For a general pair inflation rule $\sigma^{(2)}$, we can define an
associated substitution matrix\[
\mathbf{M}^{(2)}=\left(\begin{array}{cccc}
\#_{aa}(w_{aa}) & \#_{aa}(w_{ab}) & \#_{aa}(w_{ba}) & \#_{aa}(w_{bb})\\
\#_{ab}(w_{aa}) & \#_{ab}(w_{ab}) & \#_{ab}(w_{ba}) & \#_{ab}(w_{bb})\\
\#_{ba}(w_{aa}) & \#_{ba}(w_{ab}) & \#_{ba}(w_{ba}) & \#_{ba}(w_{bb})\\
\#_{bb}(w_{aa}) & \#_{bb}(w_{ab}) & \#_{bb}(w_{ba}) & \#_{bb}(w_{bb})\end{array}\right),\]
where now $\#_{ab}(w_{ba})$ denotes the number of pairs $ab$ in
the word associated with the pair $ba$. The leading eigenvalues $\lambda_{1}$
and $\lambda_{2}$ of $\mathbf{M}^{(2)}$ define another wandering
exponent\[
\omega=\frac{\ln\left|\lambda_{2}\right|}{\ln\lambda_{1}},\]
which governs the fluctuations of the letter pairs, and consequently
of the $\delta_{j}$. It is essential to note that $\omega$ is in
general different from $\omega_{l}$: for the Fibonacci sequence,
for instance, we have $\omega_{l}=-1$, but $\omega=0$.%
\footnote{There are aperiodic sequences, generated by what Hermisson called
mixed substitution rules, for which a pair inflation rule cannot be
defined; however, it is still possible to investigate the fluctuations
of the $\delta_{j}$ in terms of a set of substrings with minimal
length.%
}

By an exact renormalization-group treatment, Hermisson\cite{hermisson00}
was able to build recursion relations for effective couplings, and
to show that, in agreement with the above heuristic argument, the
eigenvalues $\lambda_{i}$ of $\mathbf{M}^{(2)}$ give directly the
RG eigenvalues $y_{i}$ around the uniform fixed point of \emph{XY}
chains with aperiodic couplings,\[
y_{i}=\frac{\ln\left|\lambda_{i}\right|}{\ln\lambda_{1}},\]
while the corresponding eigenvectors yield the scaling fields. Thus,
aperiodicity is relevant in the RG sense (i.e. it moves the RG flows
away from the uniform fixed point) if the next-to-leading eigenvalue
$y_{2}=\omega$ is positive, exactly as predicted by the Harris-Luck
criterion, Eq. (\ref{eq:harrisluck}), with $d=\nu=1$. %
\footnote{To be precise, the eigenvalue of $\mathbf{M}^{(2)}$ entering the
definition of $\omega$ is not always the next-to-leading one, but
rather the second-largest eigenvalue whose corresponding scaling field
is nonzero for a generic choice of coupling constants.%
}

For a large class of aperiodic sequences fulfilling Eq. (\ref{eq:critcond}),
$\lambda_{2}$ is given in the \emph{XX} limit by an integer $k$.
When a pair substitution rule can be defined, this integer is simply
given by\[
k=\#_{ab}(w_{ab})-\#_{ba}(w_{ab}).\]
 Thus, the wandering exponent in the \emph{XX} limit is of the form\begin{equation}
\omega=\frac{\ln k}{\ln\tau},\label{eq:omegaxx}\end{equation}
where $\tau\equiv\lambda_{1}$ corresponds to the rescaling factor
of the sequence of letter pairs.

It is also possible to determine the scaling of the lowest energy
levels $\Lambda$ with the system size $r$. For irrelevant or marginal
aperiodicity ($\omega\leq0$) we have\begin{equation}
\Lambda\sim r^{-z},\label{eq:xxmargscal}\end{equation}
with a dynamical exponent $z$ equal to unity if $\omega<0$, but
which can vary with the coupling ratio $J_{a}/J_{b}$ in the marginal
cases ($\omega=0$). Relevant aperiodicity ($\omega>0$) leads to
a different scaling form,\begin{equation}
\Lambda\sim\exp\left(-\mu r^{\omega}\right),\qquad(\mu=\textrm{constant})\label{eq:xxrelscal}\end{equation}
and to a formally infinite dynamical exponent. From Eqs. (\ref{eq:xxmargscal})
and (\ref{eq:xxrelscal}), scaling forms for low-temperature thermodynamic
properties such as the specific heat and zero-field susceptibility
can be obtained, as discussed in the next sections. Ground-state correlation
functions, however, do not seem to be simply accessible from the exact
RG treatment.

Since the critical phase of uniform \emph{XXZ} chains is also unstable
towards dimerization in the whole anisotropy regime $0<\Delta\leq1$,
one might expect that the relevant geometrical fluctuations in the
presence of aperiodic couplings would be somehow related to the $\delta_{j}$
defined above.%
\footnote{A precise definition of the relevant local parameters would require
a generalization of the criticality condition {[}Eq. (\ref{eq:critcond}){]}
to \emph{XXZ} chains, which, to the best of our knowledge, is not
currently available.%
} Consequently, the exponent $\omega$ would be involved in determining
the scaling behavior of thermodynamic properties of aperiodic chains
for all anisotropies intermediate between the \emph{XX} and Heisenberg
limits. The results of the next sections indeed provide evidence that
this seems to be the case.

\section{The Ma-Dasgupta-Hu method for aperiodic \emph{XXZ} chains\label{sec:mdhxxz}}

We now wish to investigate the effects of aperiodic couplings on \emph{XXZ}
chains described by the Hamiltonian in Eq. (\ref{eq:xxzhamilt}).
Based on the success of the Ma-Dasgupta-Hu scheme in predicting the
properties of random-bond chains, we expect that it also works in
the presence of aperiodicity. We concentrate on the case of uniform
anisotropy ($\Delta_{i}\equiv\Delta$), but more general situations
can be considered.

Applying the MDH method to aperiodic chains requires taking into account
that now, since we have only two distinct coupling constants, there
are many spin blocks with the same (largest) gap at a given energy
scale. Also, those blocks may consist of more than two spins, in which
case effective spins would form upon renormalization. The strategy
is to sweep through the lattice until all blocks with the same gap
have been renormalized, leading to new effective couplings (and possibly
spins). Then we search for the next largest gap, which again corresponds
to many blocks. When all possible original blocks have been considered,
there remains some unrenormalized spins, possibly along with effective
ones, defining new blocks which form a second generation of the lattice.
The process is then iterated, leading to the renormalization of the
spatial distribution of effective blocks (or bonds) along the generations. 

Due to the self-similarity inherent to aperiodic sequences generated
by inflation rules, it is natural that the block distribution reaches
a periodic attractor (usually a fixed point or a two-cycle) after
a few lattice sweeps; numerical implementations of the method indicate
that this attractor is independent of the anisotropy $\Delta$ for
all coupling ratios. By studying recursion relations for the effective
couplings, we can obtain analytical results. As the RG steps proceed,
the coupling ratio usually gets smaller, suggesting that the method
becomes asymptotically exact. This picture holds for marginal ($\omega=0$)
and relevant ($\omega>0$) aperiodicity. Irrelevant aperiodicity is
characterized by a wandering exponent $\omega<0$, meaning that geometric
fluctuations become negligible at long distances. An example is provided
by the Thue-Morse sequence, generated by the substitution rule $a\to ab$,
$b\to ba$, for which $\omega=-\infty$. Applying the MDH scheme to
the Thue-Morse sequence leads to an effective coupling ratio which
approaches unity along the generations, although the couplings themselves
become smaller. This means that the perturbative approach in the core
of the MDH scheme eventually breaks down, and no asymptotic behavior
can be obtained. However, this intuitively agrees with the picture
that irrelevant aperiodicity leads to the same critical properties
as the uniform model, where all couplings have the same value.

For sequences where the fraction of letters $a$ (or $b$) at odd
bonds is different from that at even bonds (i.e. where the sequence
induces average dimerization), one generally expects that a finite
gap opens between the global ground state and the first excited states,
independent of the value of the wandering exponent $\omega$. This
is the case of the period-doubling sequence, built from the substitution
rule $a\to ab$, $b\to aa$. Upon application of the MDH method, after
a few lattice sweeps (with the precise number depending on the strength
of dimerization) we reach a situation where, say, all strong bonds
occupy even positions, whereas all bonds at odd positions are weaker.
Thus, all remaining couplings are necessarily decimated in a last
lattice sweep, generating a final effective coupling which approaches
zero exponentially with the system size $N$. This can be interpreted
as indicating that there is no correlation between spins separated
by large distances, in agreement to what happens in gapped Heisenberg
and \emph{XX} chains. In the presence of average dimerization, few
quantitative predictions can be drawn from the MDH method; one of
them is an estimate of the excitation gap, whose order of magnitude
is provided by the value of the strong bonds in the final lattice
sweep. In contrast, randomly dimerized Heisenberg chains in the strong-randomness
limit are in a gapless Griffiths phase, exhibiting short-range correlations
but a diverging susceptibility.\cite{hyman96}

In a general situation, the blocks to be renormalized consist of $n$
spins connected by equal bonds $J_{0}$, and coupled to the rest of
the chain through weaker bonds $J_{l}$ and $J_{r}$. As discussed
in Appendix \ref{sec:multi}, the ground state for blocks with an
even number of spins is a singlet (as in the original MDH method),
and at low energies we can eliminate the whole block, along with $J_{l}$
and $J_{r}$, leaving an effective antiferromagnetic bond $J^{\prime}$
coupling the two spins closer to the block and given by second-order
perturbation theory as \[
J^{\prime}=\gamma_{n}\frac{J_{l}J_{r}}{J_{0}}\qquad\textrm{($n$ even)},\]
 with $\Delta$-dependent coefficients $\gamma_{n}$. On the other
hand, a block with an odd number of spins has a doublet as its ground
state; at low energies, it can be replaced by an effective spin connected
to its nearest neighbors by antiferromagnetic effective bonds \[
J_{l,r}^{\prime}=\gamma_{n}J_{l,r}\qquad\textrm{($n$ odd)},\]
 whose values are calculated by first-order perturbation theory. 

In general, the anisotropy parameters are also renormalized and become
site-dependent; for $n$ even, the effective anisotropy is $\Delta^{\prime}=\delta_{n}(\Delta_{0})\Delta_{l}\Delta_{r}$,
while for $n$ odd $\Delta_{l,r}^{\prime}=\delta_{n}(\Delta_{0})\Delta_{l,r}$,
with $\left|\delta_{n}(\Delta)\right|<1$ for $0\leq\Delta<1$ and
$\delta_{n}(1)=1$. So, for $0<\Delta<1$ the $\Delta_{i}$ flow to
the \emph{XX} fixed point (all $\Delta_{i}=0$), ultimately reproducing
the corresponding scaling behavior, while for the Heisenberg chain
all $\Delta_{i}$ remain equal to unity. An analytical treatment of
the intermediate anisotropy regime is possible (see Ref. \onlinecite{hida04b}),
leading to a prediction of the effective coupling ratio for which
the system crosses over to the \emph{XX} behavior. However, for simplicity,
we present analytical calculations for the \emph{XX} and Heisenberg
limits, showing some numerical results for the general case $0\leq\Delta\leq1$.
If we start with a uniform anisotropy $\Delta>1$, the $\Delta_{i}$
grow without limit, and the system ultimately behaves like an antiferromagnetic
Ising chain, suppressing all quantum fluctuations. For $\Delta<0$,
the $\gamma_{n}$ coefficients for $n$ even become larger than unity,
so that, if the modulation is not strong enough, the MDH scheme may
produce effective couplings which are larger than the original couplings,
leading to `bad' decimations; moreover, the two-spin local gap closes
as $\Delta\rightarrow-1$. This puts the MDH results under suspicion,
requiring a more careful analysis which is beyond the scope of the
present work.

Correlation functions can be calculated at zeroth order by assuming
that only spins which eventually appear in the same renormalized block
are correlated. Note that an effective spin represents all spins in
an original block via Clebsch-Gordan coefficients (see Appendix \ref{sec:multi}),
and this allows us to calculate correlations between any two spins
whose effective spins end up in the same block at some stage of the
RG process. In order to estimate correlations between other spin pairs
we must expand the local ground states up to second order in $J_{l,r}/J_{0}$.
This requires lengthy calculations (see Appendix \ref{sec:corr2}),
and we restrict applications of this expansion to the simplest yet
illustrative cases of sequences where only two-spin blocks are involved
in the RG steps.

In the next two sections, we present a detailed discussion of the
results obtained by applying the MDH scheme to sequences inducing
marginal or relevant aperiodicity.

We should mention that similar strong-modulation perturbative approaches
have been applied to investigate the spectral properties of noninteracting
electrons with aperiodic hopping parameters or single-site potentials
(see e.g. Refs. \onlinecite{niu86, barache94, fujita00}). However,
the \emph{XXZ} chain with nonzero anisotropy $\Delta$ is mapped by
the Jordan-Wigner transformation onto a half-filled \emph{interacting}
electron system, for which, to the best of our knowledge, no such
studies exist.

\section{Marginal aperiodicity\label{sec:marg}}

\subsection{The Fibonacci sequence}

\begin{figure}
\includegraphics[%
  width=1.0\columnwidth]{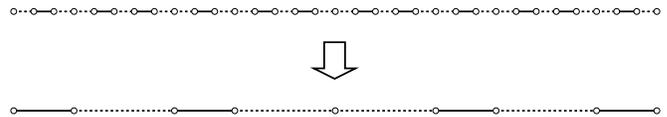}

\caption{\label{fig:fibfullab}Left end of the Fibonacci \emph{XXZ} chain
with $J_{a}<J_{b}$. Dashed (solid) lines represent weak (strong)
bonds, while circles indicate the positions of the spins. Apart from
a few bonds close to the chain ends, the effective couplings also
form a Fibonacci sequence.}
\end{figure}
First we apply the method to chains with Fibonacci couplings. This
is the simplest example of the quasiperiodic precious-mean sequences
with marginal fluctuations.\cite{hermisson00} A few bonds closer
to the left end of the original chain, along with induced effective
couplings, are shown in Fig. \ref{fig:fibfullab} for $J_{a}<J_{b}$.
In this case, only singlets are formed by the RG process; apart from
a few bonds close to the chain ends, the renormalized lattice is again
a Fibonacci chain. An effective coupling $J_{b}^{\prime}$ is induced
between spins separated by only one singlet pair, while $J_{a}^{\prime}$
connects spins separated by two singlet pairs, and in terms of the
original couplings we have\[
J_{a}^{\prime}=\gamma_{2}^{2}\frac{J_{a}^{3}}{J_{b}^{2}}\quad\text{and}\quad J_{b}^{\prime}=\gamma_{2}\frac{J_{a}^{2}}{J_{b}}.\]
 The bare coupling ratio is $\rho=J_{a}/J_{b}$, its renormalized
value being $\rho^{\prime}=\gamma_{2}\rho$. In each generation $j$,
all decimated blocks have the same size $r_{j}$ and gap $\Lambda_{j}$
(proportional to the effective $J_{b}$ bonds). The recursion relations
for $\rho$ and $\Lambda$ are given by\begin{equation}
\rho_{j+1}=\gamma_{2}\rho_{j}\quad\text{and}\quad\Lambda_{j+1}=\gamma_{2}\rho_{j}^{2}\Lambda_{j}.\label{eq:fibrr}\end{equation}
For the \emph{XX} chain $\gamma_{2}=1$, and thus $\rho_{j+1}=\rho_{j}$,
corresponding to a line of fixed points. On the other hand, for the
Heisenberg chain $\gamma_{2}=\nicefrac{1}{2}$, so that $\rho_{j+1}<\rho_{j}$,
leading to a stable fixed point $\rho_{\infty}=0$; since the perturbative
approach on which the MDH scheme is based works for $\rho\ll1$, the
method can be expected to yield asymptotically exact results for the
Heisenberg Fibonacci chains. In both cases, solving Eqs. (\ref{eq:fibrr})
gives the gap in the $j$th generation in terms of the original coupling
ratio $\rho$ and gap $\Lambda$,\[
\Lambda_{j}=\gamma_{2}^{j^{2}}\rho^{2j}\Lambda.\]
(Notice that, since $\gamma_{2}$ depends on the anisotropy, this
last equation is valid only for $\Delta_{i}\equiv0$ or $\Delta_{i}\equiv1$;
in the intermediate anisotropy regime, the variation of $\gamma_{2}$
along the generations must be taken into account.\cite{hida04b})
The distance between spins forming a singlet in the $j$th generation
defines a characteristic length $r_{j}$, corresponding to the Fibonacci
numbers $r_{j}=1$, $3$, $13$, $55$, $\ldots$; for $j\gg1$ the
ratio $r_{j+1}/r_{j}$ approaches $\phi^{3}$, where $\phi=(1+\sqrt{5})/2$
is the golden mean. So we have $r_{j}\sim r_{0}\phi^{3j}$, where
$r_{0}$ is a constant, and we obtain the dynamical scaling relation
\begin{equation}
\Lambda_{j}\sim r_{j}^{-\zeta}e^{-\mu\ln^{2}(r_{j}/r_{0})},\label{eq:gapheisfib}\end{equation}
 with $\zeta=-\nicefrac{2}{3}\ln\rho/\ln\phi$ and $\mu=-\ln\gamma_{2}/9\ln^{2}\phi$.
For the Heisenberg chain ($\gamma_{2}=\nicefrac{1}{2}$), Eq. (\ref{eq:gapheisfib})
describes a weakly exponential scaling (with a formally infinite dynamical
exponent), but not of the form $\Lambda\sim\exp\left(-r^{\omega}\right)$
found for the \emph{XX} chain with relevant aperiodicity ($\omega>0$).
For the \emph{XX} chain ($\gamma_{2}=1$), $\mu=0$ and we can identify
$\zeta$ with a dynamical exponent $z$, whose value depends on the
coupling ratio, leading to nonuniversal scaling behavior, characteristic
of strictly marginal operators. (We can check that $z=\zeta$ corresponds
to the asymptotic form of the exact \emph{XX} expression\cite{luck86,hermisson00}
for $\rho\ll1$.) This nonuniversality should hold in the anisotropy
regime $0<\Delta<1$ with a {}`bare' value of $\rho$ defined
at a crossover scale. Note that, taking into account the scaling form
$\Lambda\sim\exp\left(-r^{\omega}\right)$ valid for relevant aperiodicity,
we can view the above Heisenberg scaling form ($\mu\neq0$) as a marginally
relevant $(\omega\rightarrow0^{+}$) case. The result in Eq. (\ref{eq:gapheisfib})
has also been obtained in Ref. \onlinecite{hida04a}.

\begin{figure}
\includegraphics[%
  width=1.0\columnwidth]{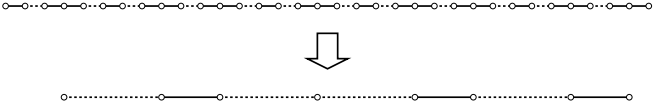}

\caption{\label{fig:fibfullba}Left end of the Fibonacci chains with $J_{a}>J_{b}$.
Effective spins form in the first lattice sweep, giving rise to a
Fibonacci chain with the roles of the weak and strong bonds interchanged,
exactly as in the original lattice in Fig. \ref{fig:fibfullab}.}
\end{figure}
If we choose $J_{a}>J_{b}$, blocks with three spins connected by
two strong bonds appear in the chain, producing effective spins upon
renormalization. However, as illustrated in Fig. \ref{fig:fibfullba},
the first lattice sweep yields again a Fibonacci chain with the roles
of weak and strong bonds interchanged, exactly as in Fig. \ref{fig:fibfullab}.
The effective couplings in the second generation are given by\begin{equation}
J_{a}^{\prime}=\gamma_{2}\gamma_{3}^{2}\frac{J_{b}^{2}}{J_{a}}\quad\textrm{and}\quad J_{b}^{\prime}=\gamma_{3}^{2}J_{b},\label{eq:fibba}\end{equation}
and the coupling ratio is now \[
\rho^{\prime}=\frac{J_{b}^{\prime}}{J_{a}^{\prime}}=\frac{1}{\gamma_{2}}\frac{J_{a}}{J_{b}}=\left(\gamma_{2}\rho\right)^{-1},\]
which is larger than one, showing that $J_{a}^{\prime}<J_{b}^{\prime}$.
Thus, we can apply the same analysis as in the case with $J_{a}<J_{b}$,
but now with `bare' couplings given by Eq. (\ref{eq:fibba}). So,
in the \emph{XX} chain, since $\gamma_{2}=1$, the MDH method predicts
scaling forms which are symmetric under $\rho\rightarrow1/\rho$,
in agreement with the exact treatment.\cite{luck86,hermisson00}

The susceptibility $\chi(T)$ can be estimated\cite{fisher94} by
assuming that, at energy scale $\Lambda_{j}\sim T$, singlet pairs
are effectively frozen, while unrenormalized spins are essentially
free, contributing Curie terms to the susceptibility. Thus, if $n_{j}\sim r_{j}^{-1}$
is the number of surviving spins in the $j$th generation, $\chi(T\sim\Lambda_{j})\sim n_{j+1}/\Lambda_{j}.$
This already gives reasonable results, as indicated by comparison
with those obtained for the \emph{XX} chain from numerical diagonalization
of finite chains,\cite{vieira04} based on the free-fermion method.\cite{lieb61}
However, a more useful approximation can be obtained by noting that,
in the $j$th generation, we can view the resulting lattice as composed
of `independent' singlets in which a pair of spins is coupled via
an \emph{XXZ} interaction with effective bond and anisotropy parameters
$J_{b}^{(j)}$ and $\Delta_{b}^{(j)}$. Since the fraction of such
singlets with respect to the number of original bonds is $(n_{j}-n_{j+1})/2$,
the free energy per site of the whole system, in the presence of an
external field $h\rightarrow0$, can be estimated as\begin{equation}
f(h,T)=\frac{1}{2}\sum_{j}\frac{n_{j}-n_{j+1}}{2}F_{\mathrm{{pair}}}\left(J_{b}^{(j)},\Delta_{b}^{(j)};h,T\right),\label{eq:fhtind}\end{equation}
where $F_{\mathrm{{pair}}}(J,\Delta;h,T)$ is the free energy of a
pair of spins interacting via the Hamiltonian\[
H_{\mathrm{{pair}}}=J\left(S_{1}^{x}S_{2}^{x}+S_{1}^{y}S_{2}^{y}+\Delta S_{1}^{z}S_{2}^{z}\right)-h(S_{1}^{z}+S_{2}^{z}).\]
Iterating the recursion relations for the effective couplings $J_{a,b}$
and $\Delta_{a,b}$, we can determine their values in each generation,
and evaluate numerically the sum in Eq. (\ref{eq:fhtind}) to obtain
the free energy. Thermodynamic properties such as the zero-field susceptibility
$\chi$ and the specific heat $c$ can be obtained by the relations\[
\chi=-\left.\frac{\partial^{2}f}{\partial h^{2}}\right|_{h=0}\textrm{\quad and}\quad c=-T\frac{\partial^{2}f}{\partial T^{2}}.\]

\begin{figure}
\includegraphics[%
  width=1.0\columnwidth]{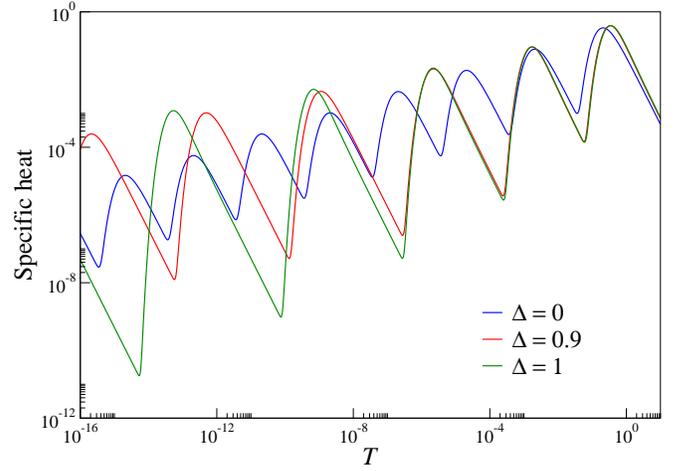}

\caption{(Color online) \label{fig:fibfullheat}Specific heat of the Fibonacci
\emph{XXZ} chains for $J_{a}/J_{b}=\nicefrac{1}{10}$ and three different
values of the uniform anisotropy $\Delta$, as given by the `independent-singlet'
approximation.}
\end{figure}
As an example, Fig. \ref{fig:fibfullheat} shows plots of the specific
heat of Fibonacci \emph{XXZ} chains with $J_{a}/J_{b}=\nicefrac{1}{10}$
and three values of the anisotropy $\Delta=\Delta_{a}=\Delta_{b}$,
corresponding to the \emph{XX} and Heisenberg limits and to an intermediate
case ($\Delta=\nicefrac{9}{10}$). The results for the \emph{XX} limit
agree very well with those obtained from numerical diagonalization,
although the agreement becomes worse for larger coupling ratios; in
particular, the specific-heat scaling law\cite{luck86,hermisson00}
\[
c(T)\sim T^{1/z}G_{c}\left(\frac{\ln T}{\ln\rho^{2}}\right),\]
with $z=\zeta(\rho)$ and $G_{c}$ a function of unit period, is fully
satisfied, reflecting the strictly marginal character of the aperiodic
perturbations. This is not the case in the Heisenberg limit, and the
logarithmic amplitudes of the oscillations in the specific heat become
larger with decreasing temperatures, reflecting the weakly exponential
dynamical scaling in Eq. (\ref{eq:gapheisfib}). For intermediate
anisotropies, there is a crossover from Heisenberg-like to \emph{XX}-like
behavior as the temperature is lowered; the larger amplitude of the
low-temperature oscillations corresponds to those of an \emph{XX}
chain with a `bare' coupling ratio $\rho_{\mathrm{{eff}}}<\rho$ defined
at a crossover scale in which effective anisotropies become negligible.
(For a detailed analysis, see Ref. \onlinecite{hida04b}.)

\begin{figure}
\includegraphics[%
  width=1.0\columnwidth]{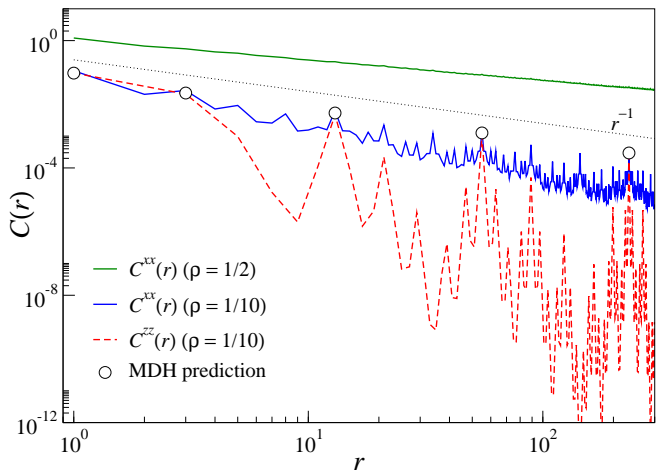}

\caption{\label{fig:fbcorr}(Color online) Ground-state correlations as a
function of the distance between spins for the Fibonacci \emph{XX}
chain. The curves are obtained from numerical diagonalization of closed
chains with $2584$ sites. For $J_{a}/J_{b}=\nicefrac{1}{10}$ (lower
curves), dominant correlations correspond to distances $r_{j}=1$,
$3$, $13$, $55$ and $233$, for which $C^{xx}$ and $C^{zz}$ are
nearly equal, as predicted by the MDH method (circles), and decay
as $1/r$ (dotted curve). Larger coupling ratios lead to a slower
decay of $C^{xx}$ (and a faster decay of $C^{zz}$), as seen for
$J_{a}/J_{b}=\nicefrac{1}{2}$ (upper curve, offset for clarity).}
\end{figure}
As all singlets formed in the $j$th generation have length $r_{j}$
and the bond distribution is fixed, the average ground-state correlation
between spins separated by a distance $r_{j}$ can be estimated as
\begin{equation}
C^{\alpha\alpha}(r_{j})\equiv\overline{\left\langle S_{i}^{\alpha}S_{i+r_{j}}^{\alpha}\right\rangle }\simeq\tfrac{1}{2}\left|c_{0}\right|\left(n_{j}-n_{j+1}\right)=\sigma\left|c_{0}\right|r_{j}^{-1},\label{eq:corrfib}\end{equation}
 where the bar denotes average over all possible pairs, $\sigma$
is a constant, $\alpha=x,y,z$, and $c_{0}$ is the correlation between
the two spins in a singlet, given by $c_{0}=-\nicefrac{1}{4}$ for
the Heisenberg chain and for both $\alpha=x$ and $\alpha=z$ in the
\emph{XX} chain. We point out that these should be the dominant correlations,
and spins separated by distances other than $r_{j}$ are predicted
to be only weakly correlated. As shown in Fig. \ref{fig:fbcorr},
results from numerical diagonalization for the \emph{XX} Fibonacci
chain with $\rho=\nicefrac{1}{10}$ agree very well with the MDH predictions.
Note that correlations in the uniform \emph{XX} chain \cite{lieb61}
decay as $C^{xx}(r)\sim r^{-1/2}$ and $C^{zz}(r)\sim r^{-2}$, so
that dominant $xx$ ($zz$) correlations in the Fibonacci chain are
weaker (stronger) than in the uniform chain. Due to the strictly marginal
character of the fluctuations induced by the aperiodic couplings,
deviations from the predictions in Eq. (\ref{eq:corrfib}) appear
in the \emph{XX} chain for larger values of $\rho$, as also shown
in the figure. This point will be further discussed in the next subsection,
but these deviations should not be present in the Fibonacci Heisenberg
chain, where aperiodicity can be viewed as marginally relevant.

\subsection{The silver-mean sequence}

\begin{figure}
\includegraphics[%
  width=1.0\columnwidth]{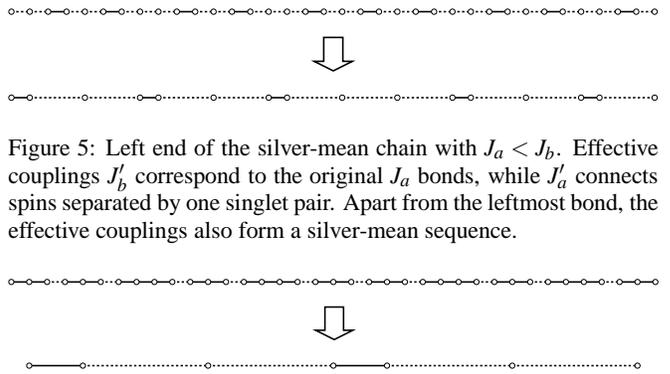}

\caption{\label{fig:smfullab}Left end of the silver-mean chain with $J_{a}<J_{b}$.
Effective couplings $J_{b}^{\prime}$ correspond to the original $J_{a}$
bonds, while $J_{a}^{\prime}$ connects spins separated by one singlet
pair. Apart from the leftmost bond, the effective couplings also form
a silver-mean sequence.}
\end{figure}
\begin{figure}
\includegraphics[%
  width=1.0\columnwidth]{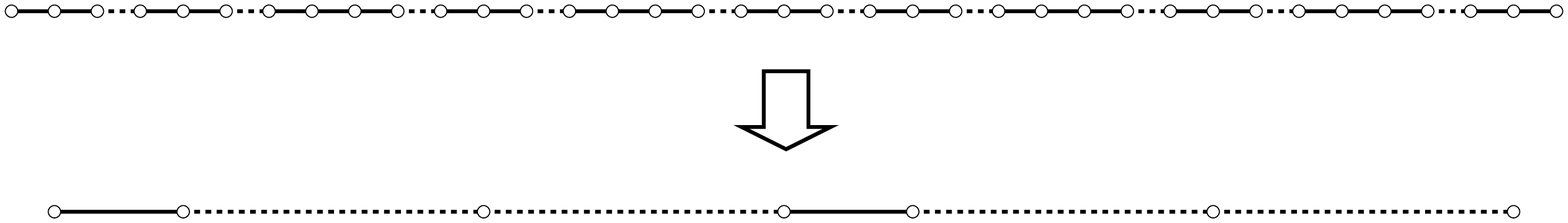}

\caption{\label{fig:smfullba}Left end of the silver-mean chain with $J_{a}>J_{b}$.
Effective spins form in the first lattice sweep, producing the same
structure as the third generation for $J_{a}<J_{b}$.}
\end{figure}
The silver-mean sequence is obtained from the substitution rule $a\rightarrow aab$,
$b\rightarrow a$, and the rescaling factor predicted in the \emph{XX}
limit is \cite{hermisson00}\[
\tau_{\mathrm{{sm}}}=1+\sqrt{2}.\]
When the MDH scheme is applied, the first lattice sweep also generates
a silver-mean sequence, identical to the original one for $J_{a}<J_{b}$,
but with the roles of weak and strong bonds interchanged for $J_{a}>J_{b}$,
as shown in Figs. \ref{fig:smfullab} and \ref{fig:smfullba}. In
the latter case, the second-generation structure is identical to the
third-generation lattice obtained for $J_{a}<J_{b}$, a situation
we can assume without loss of generality. So, we can write the recursion
relations\[
J_{a}^{\prime}=\gamma_{2}\frac{J_{a}^{2}}{J_{b}}\quad\textrm{and}\quad J_{b}^{\prime}=J_{a},\]
from which we get\[
\rho^{\prime}\equiv\frac{J_{a}^{\prime}}{J_{b}^{\prime}}=\gamma_{2}\frac{J_{a}}{J_{b}}\equiv\gamma_{2}\rho\quad\textrm{and}\quad\frac{\Lambda^{\prime}}{\Lambda}=\frac{J_{b}^{\prime}}{J_{b}}=\rho.\]
These are similar to the relations found for the Fibonacci chains.
The length of singlets formed in the $j$th generation is $r_{j}=1,$
$1$, $3$, $7$, $17$, $41$, $\ldots$, whose asymptotic ratio
is $r_{j+1}/r_{j}=\tau_{\mathrm{{sm}}}.$ Thus, solving the recursion
relations yields\[
\Lambda_{j}\sim r_{j}^{-\zeta(\rho)}e^{-\mu\ln^{2}\left(r_{j}/r_{0}\right)},\]
with \begin{equation}
\zeta(\rho)=-\frac{\ln\left(\gamma_{2}^{-\nicefrac{1}{2}}\rho\right)}{\ln\tau_{\mathrm{{sm}}}}\quad\textrm{and}\quad\mu=-\frac{\ln\left(\gamma_{2}^{\nicefrac{1}{2}}\right)}{\ln^{2}\tau_{\mathrm{{sm}}}},\label{eq:smzetamu}\end{equation}
so that in the \emph{XX} limit the scaling again corresponds to a
nonuniversal power-law behavior with a dynamical exponent $z=\zeta(\rho).$

\begin{figure}
\includegraphics[%
  width=1.0\columnwidth]{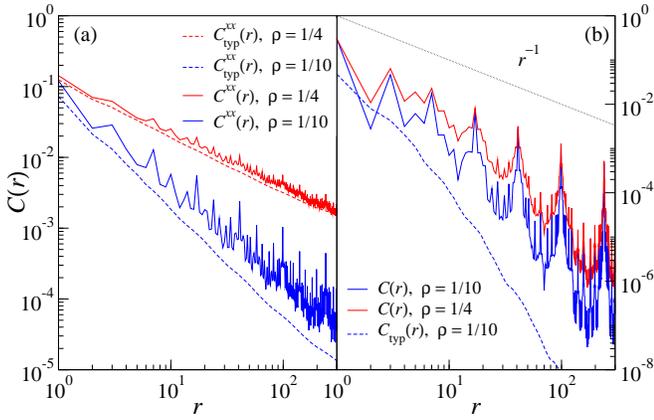}

\caption{\label{fig:smcorr}(Color online) Ground-state pair correlations
in the silver-mean \emph{XX} (a) and Heisenberg (b) chains, for $\rho=\nicefrac{1}{4}$
(upper curves) and $\rho=\nicefrac{1}{10}$ (lower curves), obtained
from the second-order MDH scheme (chains with $8119$ sites). Solid
and dashed curves correspond to average and typical correlations,
respectively.}
\end{figure}
As in the Fibonacci chains, pair correlations in the ground state
can be estimated by noting that only singlets are produced by the
RG process, and we conclude that for $\rho\ll1$ the dominant correlations
(those between spins separated by the characteristic distances $r_{j}=1$,
$3$, $7$, $17$,$\ldots$) should behave as\[
C(r_{j})\sim\frac{1}{r_{j}},\]
while correlations between spins separated by other distances should
be negligible. 

However, for \emph{XX} chains, this is a rough approximation if the
coupling ratio is not too small, and free-fermion calculations reveal
a power-law decay of both $C^{xx}(r_{j})$ and $C^{zz}(r_{j})$ with
$\rho$-dependent exponents, as in Fig. \ref{fig:fbcorr}. This can
be accounted for by the MDH method if we expand the ground-state vector
to second-order in $\rho$, as described in Appendix \ref{sec:corr2}.
Results of such calculations are shown in Fig. \ref{fig:smcorr} for
$\rho=\nicefrac{1}{4}$ and $\rho=\nicefrac{1}{10}$, in the \emph{XX}
and Heisenberg limits. Both average and typical correlations are plotted;
the latter, defined by\[
C_{\mathrm{typ}}^{\alpha\alpha}(r)=\exp\left(\overline{\ln\left|\left\langle S_{i}^{\alpha}S_{i+r}^{\alpha}\right\rangle \right|}\right),\]
 filter out the contribution of those pairs of spins most strongly
correlated, yielding an estimate of the correlation between two arbitrary
spins separated by a distance $r$. In the random-singlet phase, characteristic
of random-bond chains,\cite{fisher94} average correlations decay
algebraically as $C(r)\sim1/r^{2}$, whereas typical correlations
are short-ranged, following $C_{\mathrm{typ}}(r)\sim\exp(-\sqrt{r/r_{0}})$.
This is due to the fact that average correlations are dominated by
the rare singlet pairs, while the correlation between a typical pair
of spins is of the order of some intermediate effective coupling (see
Appendix \ref{sec:corr2}). 

As shown in Fig. \ref{fig:smcorr}(a), this picture does not hold
for silver-mean \emph{XX} spin chains. As the coupling ratio is lowered,
average and typical correlations exhibit clearly distinct behavior,
but both $C^{xx}(r_{j})$ and $C_{\mathrm{typ}}^{xx}(r)$ still follow
approximately a power law, with $\rho$-dependent exponents, reproducing
the results of the free-fermion calculations. This nonuniversality
is related to the marginal character of the precious-mean fluctuations,
which keeps the effective coupling ratio unchanged along the RG process,
and can be qualitatively understood from the following argument. For
each singlet pair coupled by a strong bond and whose spins are separated
by a characteristic distance $r_{j}$, there exists a certain number
of other spin pairs separated by the same distance $r_{j}$, but connected
through weaker bonds, whose correlation (see Appendix \ref{sec:corr2})
is smaller than the strongest ones by factors of order $\rho$, $\rho^{2}$,
$\rho^{3}$, etc. The average correlation can be estimated as\begin{equation}
C^{xx}(r_{j})\sim\left(1+a_{1}\rho+a_{2}\rho^{2}+\cdots+a_{j}\rho^{j}\right)r_{j}^{-1},\label{eq:corrhier}\end{equation}
where the $a_{n}$'s are proportional to the fractions of pairs giving
contributions of order $\rho^{n}$, and the sum has an upper cutoff
at $n=j$, since $r_{j}$ corresponds to the $j$th generation singlets.
Assuming that $a_{n}=a_{1}\alpha^{n-1}$, for some constant $\alpha$
(which can be numerically checked to be a reasonable approximation
for small $\rho$), we have\[
1+a_{1}\rho+a_{2}\rho^{2}+\cdots+a_{j}\rho^{j}=1+\frac{1-(\alpha\rho)^{j}}{1-\alpha\rho}a_{1}\rho,\]
and taking into account that $r_{j}\sim r_{0}\tau_{\mathrm{sm}}^{j}$
we can write\[
(\alpha\rho)^{j}\sim r_{j}^{1-\eta(\rho)},\quad\textrm{with}\quad\eta(\rho)=1-\frac{\ln(\alpha\rho)}{\ln\tau_{\mathrm{sm}}}.\]
 Combining the above results we conclude that \[
C^{xx}(r_{j})\sim\frac{1}{r_{j}},\]
for $\rho<\alpha^{-1}$, reproducing the zeroth order MDH prediction,
but a nonuniversal behavior\[
C^{xx}(r_{j})\sim\frac{1}{r_{j}^{\eta(\rho)}}\]
is obtained for $\rho>\alpha^{-1}$. 

\begin{figure}
\includegraphics[%
  width=0.88\columnwidth]{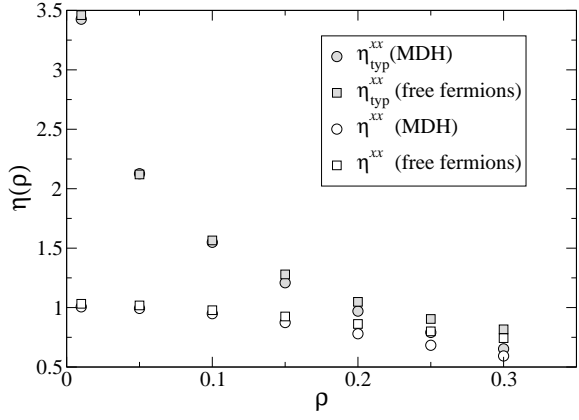}

\caption{\label{fig:smexp}Decay exponents of $C^{xx}(r_{j})$ and $C_{\mathrm{typ}}^{xx}(r)$
for the \emph{XX} silver-mean chain as a function of the coupling
ratio $\rho=J_{a}/J_{b}$, obtained from both the second-order MDH
scheme (chains with 8819 sites) and the free-fermion method (chains
with 3363 sites). Errors bars are at most the same size as the symbols
themselves.}
\end{figure}
For the silver-mean \emph{XX} chains, an estimate of the $a_{n}$
based on numerical implementations of the MDH method gives $a_{1}\simeq9.5$
and $\alpha\simeq3$, but with some dependence on $r_{j}$ and $\rho$.
As shown in Fig. \ref{fig:smexp}, the decay exponent $\eta^{xx}$
of the average correlations approaches unity as $\rho\rightarrow0$,
but starts to decrease more rapidly for $\rho\gtrsim0.1$, considerably
less than $1/\alpha$; this discrepancy indicates that Eq. (\ref{eq:corrhier}),
with the assumption of constant $a_{n}$'s, although providing a valuable
insight into the origin of the nonuniversal behavior, is not a good
approximation for larger coupling ratios. The exponents predicted
by the second-order MDH scheme are systematically smaller than the
presumably exact ones obtained from the free-fermion method (which
should tend to $\eta^{xx}=\nicefrac{1}{2}$ as $\rho\rightarrow1$).
This also happens for the decay exponent $\eta_{\mathrm{typ}}^{xx}$
of the typical correlations, which diverges as $\rho\rightarrow0$,
in agreement with the fact that, in this limit, the chain decomposes
into independent singlets. A similar behavior is observed for the
transverse correlations $C^{zz}(r)$ and $C_{\mathrm{{typ}}}^{zz}(r)$
(but now the decay exponents approach $\eta^{zz}=\eta_{\mathrm{{typ}}}^{zz}=2$
as $\rho\rightarrow1$).

On the other hand, dominant ground-state correlations in the Heisenberg
silver-mean chain closely follow the predictions of the zeroth-order
MDH scheme, as can be seen in Fig. \ref{fig:smcorr}(b). This is due
to the fact that the effective coupling ratio decreases as the RG
proceeds, and the contribution to $C(r_{j})$ due to spin pairs other
than those connected by strong bonds becomes exponentially negligible.
Typical correlations decay not as a power law, but rather according
to \[
C_{\mathrm{typ}}(r)\sim e^{-\mu_{\mathrm{{typ}}}\ln^{2}(r/r_{0})},\]
precisely the same form of the dynamical scaling; by fitting the numerical
results, the constant $\mu_{\mathrm{{typ}}}$ is found to be approximately
$\mu/2$, with $\mu$ given by Eq. (\ref{eq:smzetamu}). As in the
random-singlet phase, the scaling form of the typical correlations
is similar to that of the lowest gaps, reflecting the fact that two
spins separated by a distance $r$ are basically uncorrelated until
the energy scale is of order $\Lambda(r)$, when they become weakly
correlated through an intervening spin taking part in a singlet pair.

\subsection{\label{sec:bm}The bronze-mean sequence}

\begin{figure}
\includegraphics[%
  width=1.0\columnwidth]{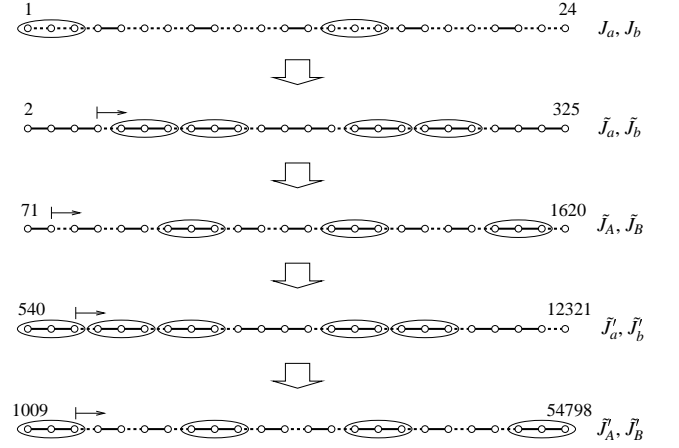}

\caption{\label{fig:bmfullab}First five generations of the bronze-mean \emph{XXZ}
chain with $J_{a}<J_{b}$, each showing the leftmost $24$ sites.
The numbers indicate the positions of the sites in the original chain.
Encircled blocks contribute effective spins when renormalized. The
labels on the right denote the effective bonds in each generation.
The attractor of the block distribution is a two-cycle, reached at
the second generation. All bonds to the right of the horizontal arrow
follow the same sequence in the second (third) and fourth (fifth)
generations.}
\end{figure}
 The bronze-mean sequence is built from the substitution rule $a\rightarrow aaab$,
$b\rightarrow a$, with a large \emph{XX} rescaling factor \cite{hermisson00}\[
\tau_{\mathrm{{bm}}}=\left(\frac{3+\sqrt{13}}{2}\right)^{3}\simeq36.03.\]
The bond-distribution attractor produced by the MDH RG scheme is not
a fixed point, but a two-cycle;%
\footnote{The presence of two-cycles in association with aperiodicity is not
uncommon. In ferromagnetic $q$-state Potts models with a suitable
choice of aperiodic couplings, for sufficiently large $q$ the finite-temperature
critical behavior is governed by a two-cycle, the uniform fixed point
being unstable; see e.g. T. A. S. Haddad, S. T. R. Pinho, and S. R.
Salinas, Phys. Rev. E \textbf{61}, 3330 (2000).%
} apart from a few bonds near the chain ends, the same distributions
alternate between even and odd generations, as shown for $J_{a}<J_{b}$
in Fig. \ref{fig:bmfullab}. In this case, the second-generation couplings
relate to the original couplings by\[
\tilde{J}_{a}=\gamma_{2}^{7}\gamma_{3}^{2}\frac{J_{a}^{5}}{J_{b}^{4}}\quad\textrm{and\quad}\tilde{J}_{b}=\gamma_{2}^{5}\gamma_{3}^{2}\frac{J_{a}^{4}}{J_{b}^{3}}.\]
Likewise, in terms of the couplings in the previous generation, we
write the third-generation couplings,\[
\tilde{J}_{A}=\gamma_{3}^{2}\gamma_{4}\frac{\tilde{J}_{a}^{2}}{\tilde{J}_{b}}\quad\textrm{and\quad}\tilde{J}_{B}=\gamma_{3}^{2}\tilde{J}_{a},\]
and the fourth-generation couplings,\[
\tilde{J}_{a}^{\prime}=\gamma_{2}^{3}\gamma_{3}^{2}\frac{\tilde{J}_{A}^{4}}{\tilde{J}_{B}^{3}}\quad\textrm{and\quad}\tilde{J}_{b}^{\prime}=\gamma_{2}^{2}\gamma_{3}^{2}\frac{\tilde{J}_{A}^{3}}{\tilde{J}_{B}^{2}}.\]
Since the attractor is now a two-cycle, and not a fixed point, we
must relate the couplings in the fourth and second-generations. By
eliminating $\tilde{J}_{A}$ and $\tilde{J}_{B}$ in the above equations,
we get \[
\tilde{J}_{a}^{\prime}=\gamma_{2}^{3}\gamma_{3}^{4}\gamma_{4}^{4}\frac{\tilde{J}_{a}^{5}}{\tilde{J}_{b}^{4}}\quad\textrm{and\quad}\tilde{J}_{b}^{\prime}=\gamma_{2}^{2}\gamma_{3}^{4}\gamma_{4}^{3}\frac{\tilde{J}_{a}^{4}}{\tilde{J}_{b}^{3}},\]
so that the coupling ratios satisfy the recursion relation\[
\tilde{\rho}^{\prime}\equiv\frac{\tilde{J}_{a}^{\prime}}{\tilde{J}_{b}^{\prime}}=\gamma_{2}\gamma_{4}\frac{\tilde{J}_{a}}{\tilde{J}_{b}}\equiv\gamma_{2}\gamma_{4}\tilde{\rho},\]
while the corresponding gaps are related by\[
\frac{\Lambda^{\prime}}{\Lambda}=\frac{\tilde{J}_{b}^{\prime}}{\tilde{J}_{b}}=\gamma_{2}^{2}\gamma_{3}^{4}\gamma_{2}^{3}\tilde{\rho}^{4}.\]
The distance between spins connected by strong bonds in the $j$th
generation is $r_{j}=1$, $13$, $43$, $469$, $1549$,$\ldots$,
which asymptotically gives $r_{j+2}/r_{j}=\tau_{\mathrm{{bm}}}$,
so that $r_{2j}\sim r_{0}\tau_{\mathrm{{bm}}}^{j}$ . Thus, solving
the recursion relations for $\tilde{\rho}$ and $\Lambda$, and taking
into account that $\tilde{\rho}=\gamma_{2}^{2}\rho$, we obtain the
dynamic scaling form\begin{equation}
\Lambda_{2j}=r_{2j}^{-\zeta(\rho)}\exp\left(-\mu\ln^{2}\frac{r_{2j}}{r_{0}}\right),\label{eq:scalingbm}\end{equation}
with\[
\zeta(\rho)=-\frac{\ln\left(\gamma_{2}^{8}\gamma_{3}^{4}\gamma_{4}\rho^{4}\right)}{\ln\tau_{\mathrm{{bm}}}}\quad\textrm{and}\quad\mu=-\frac{\ln\left(\gamma_{2}^{2}\gamma_{4}\right)}{\ln^{2}\tau_{\mathrm{{bm}}}}.\]
Of course, the same form is obtained if we choose to look at the odd
generations. In the \emph{XX} limit, as $\gamma_{2}=\gamma_{4}=1$,
we again have $\mu=0$ and Eq. (\ref{eq:scalingbm}) corresponds to
a nonuniversal power-law scaling behavior, with a dynamical exponent
$z=\zeta(\rho)$; once more, as in all marginal \emph{XX} chains,
$\zeta$ equals the leading term in the $\rho\ll1$ expansion of the
exact dynamical exponent.\cite{hermisson00} As in the Fibonacci and
silver-mean chains, choosing $J_{a}>J_{b}$ leads to the same scaling
behavior, since after the first lattice sweep the bond distribution
is essentially equal to the one obtained for $J_{a}<J_{b}$.

\begin{figure}
\includegraphics[%
  width=1.0\columnwidth]{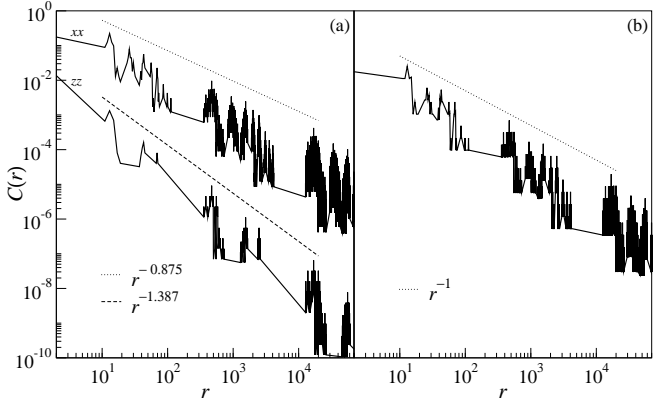}

\caption{\label{fig:bmfullcorr}Ground-state correlation functions of the
\emph{XX} (a) and Heisenberg (b) bronze-mean chains, obtained from
the zeroth-order MDH scheme (chains with $2\,010\,601$ sites). }
\end{figure}
Thus, the bronze-mean chains present qualitatively the same low-temperature
thermodynamic behavior as the Fibonacci chain. However, this is not
the case for ground-state properties. As indicated in Fig. \ref{fig:bmfullab},
the renormalization process in each generation involves all spins
in the chain, and gives rise to a hierarchy of effective spins, analogous
to that shown in Fig. \ref{fig:estfullab}. As a consequence, an effective
spin in the $j$th generation represents $3^{j-1}$ real spins. So,
while the ground states of the Fibonacci and silver-mean chains could
be described as `aperiodic singlet phases', from which excitations
of a given energy involve spins separated by a single, well-defined
distance, low-energy excitations in the bronze-mean chain involve
an exponentially large number of spin pairs, whose distances are distributed
in an increasing range. This is reflected in the ground-state correlation
functions, which exhibit a fractal-like structure, as seen in Fig.
\ref{fig:bmfullcorr}. The strongest correlations in the chains correspond
to the distances $r_{2j}=13$, $469$, $16897$, $\ldots$, and their
scaling behavior can be obtained by the following analysis. 

Consider a pair of neighboring effective spins belonging to the same
block in the $j$th generation, and let $c_{0}$ be their zeroth-order
correlation. Each of these spins represents $3^{j-1}$ real spins,
so that for each such pair there are $3^{j-1}$ pairs of real spins
separated by the same distance $r_{j}$ contributing to the total
correlation per site $C(r_{j})$. However, the contribution of a real
pair to $C(r_{j})$ depends on the string of Clebsch-Gordan coefficients
indicating the weight of its two spins in the effective spins: each
time the intermediate effective spin representing a real spin $S_{k}$
is located at the ends (the center) of a three-spin block, the weight
of $S_{k}$ is multiplied by a factor $c_{1,3}$ ($c_{2,3}$) upon
renormalization. (These coefficients are in general different for
$xx$ and $zz$ correlations; see Appendix \ref{sec:multi}.) Since
each effective spin in the $j$th generation has gone through $j-1$
renormalizations, a real pair can be classified according to the number
$n$ of factors $c_{1,3}$ present in the (equal) weights of its spins.
The contribution of all type-$n$ pairs to $C(r_{j})$ is proportional
to the number of such pairs, being given by\[
g_{j}^{(n)}=2^{n}\frac{(j-1)!}{n!(j-1-n)!}\left(c_{1,3}^{2}\right)^{n}\left(c_{2,3}^{2}\right)^{j-1-n}\left|c_{0}\right|.\]
 Thus, the total contribution of a single effective-spin pair to $C(r_{j})$
is \[
g_{j}=\sum_{n=0}^{j-1}g_{j}^{(n)}=\left(c_{2,3}^{2}+2c_{1,3}^{2}\right)^{j-1}\left|c_{0}\right|,\]
which gives \[
C(r_{j})\sim n_{j}g_{j}\sim\frac{\left(c_{2,3}^{2}+2c_{1,3}^{2}\right)^{j}}{r_{j}},\]
where $n_{j}$ is the fraction of active spins in the $j$th generation.
Since asymptotically we have $j=\ln(r_{j}/r_{0})/\ln\sqrt{\tau_{\mathrm{bm}}}$,
this last result can be written as\begin{equation}
C(r_{j})\sim r_{j}^{-\eta},\quad\textrm{with}\quad\eta=1-\frac{\ln\left(c_{2,3}^{2}+2c_{1,3}^{2}\right)}{\ln\sqrt{\tau_{\mathrm{bm}}}}.\label{eq:bmcorr}\end{equation}
For the Heisenberg chain, $c_{1,3}=\gamma_{3}=\nicefrac{2}{3}$ and
$c_{2,3}=\nicefrac{1}{3}$, so that $\eta=1$. For the \emph{XX} chain,
$\eta$ depends on whether we look at longitudinal or transverse correlations:
in the former case we have $c_{1,3}^{xx}=\nicefrac{1}{\sqrt{2}}$
and $c_{2,3}^{xx}=\nicefrac{1}{2}$, so that $\eta^{xx}\simeq0.875$,
while in the latter case $c_{1,3}^{zz}=\nicefrac{1}{2}$, $c_{2,3}^{zz}=0$,
and so $\eta^{zz}\simeq1.387$. These values are fully compatible
with the results from numerical implementations of the MDH scheme
shown in Fig. \ref{fig:bmfullcorr}, and agree very well with free-fermion
calculations for \emph{XX} chains with $\rho\ll1$. Again, larger
coupling ratios lead to nonuniversal decay of the correlations, except
in the Heisenberg limit.

\subsection{A sequence producing effective-spin triples\label{sub:est}}

\begin{figure}
\includegraphics[%
  width=1.0\columnwidth]{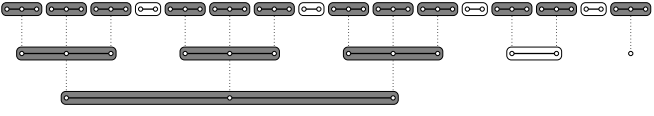}

\caption{\label{fig:estfullab}First three generations of the \emph{XXZ} chain
with couplings following the sequence in Sec. \ref{sub:est}, for
$J_{a}<J_{b}$, showing an effective-spin hierarchy. Solid lines indicate
strong bonds; for clarity, weak bonds are not represented. Shaded
blocks contribute effective spins when renormalized, while white blocks
form singlets. A third-generation effective spin represents three
second-generation and nine first-generation spins.}
\end{figure}
The appearance of an effective-spin hierarchy is better illustrated
by the sequence obtained from the substitution rule $a\rightarrow bbaba$,
$b\rightarrow bba$. The first three generations of the chains, for
$J_{a}<J_{b}$, are shown in Fig. \ref{fig:estfullab}; for $J_{a}>J_{b}$
the first lattice sweep interchanges the roles of weak and strong
bonds, recovering the former case. Renormalization involves both three-spin
and two-spin blocks, and each lattice sweep reproduces the original
sequence, yielding effective couplings given by\[
J_{a}^{\prime}=\gamma_{2}\gamma_{3}^{2}\frac{J_{a}^{2}}{J_{b}}\quad\textrm{and}\quad J_{b}^{\prime}=\gamma_{3}^{2}J_{a},\]
so that the recursion relations for the coupling ratio and the energy
gap are\[
\rho^{\prime}=\gamma_{2}\rho\quad\textrm{and}\quad\Lambda^{\prime}=\gamma_{3}^{2}\rho\Lambda.\]
The size of three-spin blocks follows $r_{j}=2$, $6$, $22$,$\ldots$,
while that of two-spin blocks corresponds to $r_{j}/2$, leading asymptotically
to a rescaling factor\[
\tau_{\mathrm{st}}=2+\sqrt{3}\simeq3.73,\]
and a dynamical scaling relation\[
\Lambda_{j}\sim r_{j}^{-\zeta(\rho)}e^{-\mu\ln^{2}\left(r_{j}/r_{0}\right)},\]
with \[
\zeta(\rho)=-\frac{\ln\left(\gamma_{2}^{-\nicefrac{1}{2}}\gamma_{3}^{2}\rho\right)}{\ln\tau_{\mathrm{{st}}}}\quad\textrm{and}\quad\mu=-\frac{\ln\left(\gamma_{2}^{\nicefrac{1}{2}}\right)}{\ln^{2}\tau_{\mathrm{st}}}.\]
Again, aperiodicity induces nonuniversal behavior for $0\leq\Delta<1$,
and a weakly exponential scaling in the Heisenberg limit.

As in the bronze-mean chains, discussed in the previous subsection,
excitations of a given energy involve an exponentially large number
of spins, due to the effective-spin hierarchy. More precisely, since
each effective spin in $j$th generation represents $3^{j-1}$ real
spins, excitations with energy $\Lambda_{j}$, corresponding to breaking
a $j$th generation singlet, involve $2\cdot3^{j-1}$ spins; exciting
a three-spin block in the same generation costs an energy of the same
order, and involves $3^{j}$ spins. Dominant ground-state correlation
functions also decay as in Eq. (\ref{eq:bmcorr}), but now with\[
\eta=1-\frac{\ln\left(c_{2,3}^{2}+2c_{1,3}^{2}\right)}{\ln\tau_{\mathrm{st}}},\]
yielding $\eta^{xx}\simeq0.830$ and $\eta^{zz}\simeq1.526$ for the
\emph{XX} chain, and $\eta=1$ for the Heisenberg chain. These values
are again fully compatible with results from numerical implementations
of the MDH scheme.

\subsection{A marginal tripling sequence\label{sub:smt}}

\begin{figure}
\includegraphics[%
  width=1.0\columnwidth]{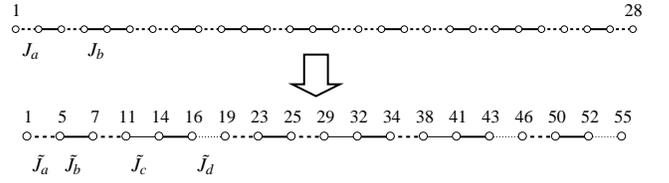}

\caption{\label{fig:smtfullab}First two generations of the \emph{XXZ} chain
with couplings following the sequence in Sec. \ref{sub:smt}, for
$J_{a}<J_{b}$. The numbers indicate the position of the spins in
the original chain. The first lattice sweep generates a fixed-point
bond distribution with four different effective couplings $\tilde{J}_{a}$
through $\tilde{J}_{d}$. }
\end{figure}
This sequence is generated by the substitution rule $a\rightarrow aba$,
$b\rightarrow bba$. As discussed in Ref. \onlinecite{hermisson00},
this type of aperiodicity may lead to marginal behavior even in anisotropic
\emph{XY} chains. As shown in Fig. \ref{fig:smtfullab}, for $J_{a}<J_{b}$
the MDH scheme produces a second-generation lattice with four different
effective couplings, given by\[
\tilde{J}_{a}=\gamma_{2}\gamma_{3}\frac{J_{a}^{2}}{J_{b}},\quad\tilde{J}_{b}=\gamma_{3}J_{a},\]
\[
\tilde{J}_{c}=\gamma_{3}^{2}J_{a},\quad\tilde{J}_{d}=\gamma_{2}\frac{J_{a}^{2}}{J_{b}},\]
with an effective coupling ratio $\tilde{\rho}=\tilde{J}_{a}/\tilde{J}_{b}=\gamma_{2}\rho$.
(Choosing $J_{a}>J_{b}$ interchanges the roles of $\tilde{J}_{c}$
and $\tilde{J}_{d}$, otherwise producing the same bond distribution.)
The bond distribution does not change upon further lattice sweeps,
and the effective couplings satisfy the recursion relations\[
\tilde{J}_{a}^{\prime}=\gamma_{2}\frac{\tilde{J}_{a}^{2}}{\tilde{J}_{b}},\quad\tilde{J}_{b}^{\prime}=\gamma_{2}\frac{\tilde{J}_{c}\tilde{J}_{d}}{\tilde{J}_{b}},\]
\[
\tilde{J}_{c}^{\prime}=\gamma_{2}\frac{\tilde{J}_{a}\tilde{J}_{c}}{\tilde{J}_{b}},\quad\tilde{J}_{d}^{\prime}=\gamma_{2}\frac{\tilde{J}_{a}\tilde{J}_{d}}{\tilde{J}_{b}}.\]
Noting that $\tilde{J}_{a}=\tilde{J}_{c}\tilde{J}_{d}/\tilde{J}_{b},$
we can write \[
\tilde{\rho}^{\prime}=\frac{\tilde{J}_{a}^{\prime}}{\tilde{J}_{b}^{\prime}}=\frac{\tilde{J}_{a}}{\tilde{J}_{b}}=\tilde{\rho},\]
so that aperiodicity is marginal even in the Heisenberg limit. The
recursion relation for the gaps is \[
\Lambda^{\prime}=\frac{\tilde{J}_{b}^{\prime}}{\tilde{J}_{b}}\Lambda=\gamma_{2}\tilde{\rho}\Lambda,\]
and the size of the singlets formed along the generations follows
$r_{j}=3^{j}-1$, with a rescaling factor $\tau_{\mathrm{mt}}=3$,
so that the dynamic scaling relation is given by\[
\Lambda_{j}\sim r_{j}^{-\zeta(\rho)},\]
with a nonuniversal dynamical exponent\[
z=\zeta(\rho)=-\frac{\ln\left(\gamma_{2}^{2}\rho\right)}{\ln3}.\]

\begin{figure}
\includegraphics[%
  width=1.0\columnwidth]{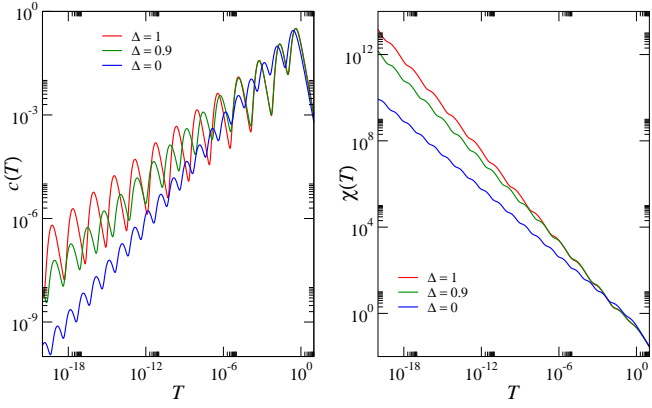}

\caption{\label{fig:mtxh}(Color online) Log-log plots of the specific heat
$c(T)$ and susceptibility $\chi(T)$ as functions of temperature
for \emph{XXZ} chains with couplings following the marginal tripling
sequence, for $J_{a}/J_{b}=\nicefrac{1}{10}$ and three different
values of the uniform anisotropy $\Delta$.}
\end{figure}
Thermodynamic properties can be estimated by using the same idea of
the `independent-singlet' approximation described for Fibonacci chains,
with slight modifications due to the fact that the first lattice sweep
(but not the later ones) involves renormalization of both two- and
three-spin blocks. Thus, the free energy per site can be calculated
by adding to Eq. (\ref{eq:fhtind}) a term representing the contribution
of spins renormalized in the first lattice sweep, and given by\begin{equation}
f_{1}(h,T)=\frac{1}{6}F_{\mathrm{pair}}(J_{b},\Delta_{b};h,T)+\frac{1}{6}F_{\mathrm{triple}}(J_{b},\Delta_{b};h,T),\label{eq:f1ht}\end{equation}
where $F_{\mathrm{triple}}(J,\Delta;h,T)$ is the free energy of a
spin triple obeying the Hamiltonian\[
\begin{array}{ccl}
H_{\mathrm{triple}} & = & J\left(S_{1}^{x}S_{2}^{x}+S_{1}^{y}S_{2}^{y}+\Delta S_{1}^{z}S_{2}^{z}\right)\\
 & + & J\left(S_{2}^{x}S_{3}^{x}+S_{2}^{y}S_{3}^{y}+\Delta S_{2}^{z}S_{3}^{z}\right)\\
 & - & h(S_{1}^{z}+S_{2}^{z}+S_{3}^{z}).\end{array}\]
Notice that three-spin blocks yield effective spins when renormalized,
and these will pair with other real or effective spins to form singlets
in the second lattice sweep, but this is not taken into account by
Eq. (\ref{eq:f1ht}). In order to obtain a correct estimate of the
low-temperature susceptibility, we must multiply the contribution
arising from spin triples by a factor like $e^{-J_{b}/T}$. Results
for the temperature dependence of the specific heat and susceptibility
are shown in Fig. \ref{fig:mtxh}, for $\rho=\nicefrac{1}{10}$ and
three values of the uniform anisotropy $\Delta$, corresponding to
the \emph{XX} and Heisenberg chains and to an intermediate case. Both
quantities exhibit log-periodic oscillations, obeying the scaling
forms\[
c(T)\sim T^{\nicefrac{1}{z}}G_{c}\left(\frac{\ln T}{\ln\lambda}\right)\quad\textrm{and}\quad\ \chi(T)\sim T^{\nicefrac{1}{z}-1}G_{\chi}\left(\frac{\ln T}{\ln\lambda}\right),\]
with $\lambda$ being the asymptotic ratio between the gaps in successive
generations, while $G_{c}$ and $G_{\chi}$ are periodic functions
(with period one). In the \emph{XX} and Heisenberg limits, we have
$\lambda=\gamma_{2}^{2}\rho$; for intermediate anisotropies, $\lambda$
equals $\rho_{\mathrm{eff}}$, a coupling ratio defined at the energy
scale in which effective anisotropies become negligible.

Also as a consequence of the strictly marginal character of aperiodic
fluctuations for all anisotropies in the regime $0\leq\Delta\leq1$,
dominant ground-state correlations follow $C(r_{j})\sim1/r_{j}$ in
the $\rho\ll1$ regime, but nonuniversal behavior should be observed
for larger coupling ratios.

\section{Relevant aperiodicity\label{sec:rel}}

\subsection{The binary Rudin-Shapiro sequence}

\begin{figure}
\includegraphics[%
  width=1.0\columnwidth]{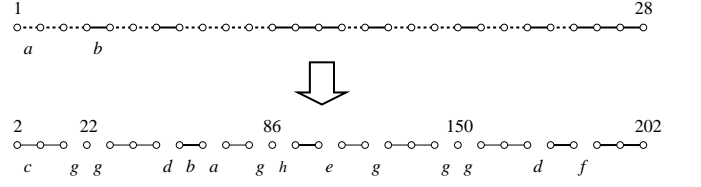}

\caption{\label{fig:rsfullab}First two generations of the binary Rudin-Shapiro
\emph{XXZ} chain, for $J_{a}<J_{b}$. The first lattice sweep generates
$8$ different effective couplings, $\tilde{J}_{a}$ through $\tilde{J}_{h}$,
labeled in the figure by the letters $a-h$. Further renormalization
does not change the bond distribution. Starting from the second generation,
only blocks coupled by $\tilde{J}_{b}$ (thick lines) and $\tilde{J}_{c}$
(thin lines) bonds are renormalized. (For clarity, weaker bonds are
not drawn in the picture.)}
\end{figure}
The Rudin-Shapiro sequence is originally defined as a four-letter
sequence,\cite{luck89} generated by the substitution rule $a\rightarrow ac$,
$b\rightarrow dc$, $c\rightarrow ab$ and $d\rightarrow db$. It
has the interesting property that its geometrical fluctuations mimic
those induced by a random distribution. In order to reduce it to a
binary sequence, we make the associations $c\equiv a$ and $d\equiv b$,
obtaining an inflation rule for letter pairs, given by $aa\rightarrow aaab$,
$ab\rightarrow aaba$, $ba\rightarrow bbab$ and $bb\rightarrow bbba$.
The rule generates blocks having between $2$ and $5$ spins, and
is symmetric under the interchange of $a$ and $b$, so that the scaling
behavior is invariant with respect to the interchange of $J_{a}$
and $J_{b}$. The left-end of the first two generations of the Rudin-Shapiro
chains is shown in Fig. \ref{fig:rsfullab} for $J_{a}<J_{b}$. 

\begin{figure}
\includegraphics[%
  width=1.0\columnwidth]{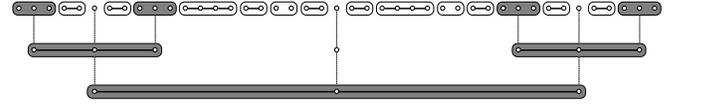}

\caption{\label{fig:rshierab}Pattern of three-spin blocks and isolated spins
leading to the effective-spin hierarchy in Rudin-Shapiro \emph{XXZ}
chains. Thick lines indicate strong bonds. Shaded blocks contribute
effective spins when renormalized; white blocks form singlets.}
\end{figure}
\begin{figure}
\includegraphics[%
  width=1.0\columnwidth]{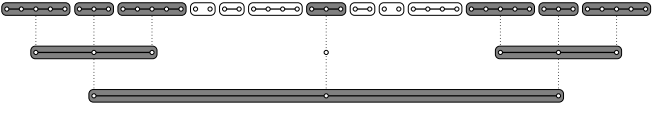}

\caption{\label{fig:rshierba}Pattern of five-spin and three-spin blocks leading
to the effective-spin hierarchy in Rudin-Shapiro \emph{XXZ} chains.
Thick lines indicate strong bonds. Shaded blocks contribute effective
spins when renormalized; white blocks form singlets.}
\end{figure}

Blocks with more than $3$ spins are eliminated in the first lattice
sweep and do not appear in later generations. Both two- and three-spin
blocks are present in the fixed-point block distribution (already
reached at the second generation), and upon renormalization the sequence
produces an effective-spin hierarchy, stemming from approximately
mirror-symmetric patterns of three-spin and five-spin blocks in the
original lattice. This is illustrated in Figs. \ref{fig:rshierab}
and \ref{fig:rshierba}.%

In the $j$th generation, three-spin blocks have size $r_{j}=2\cdot4^{j-1}$
(with a rescaling factor $\tau_{\mathrm{rs}}=4$), while two-spin
blocks have size $r_{j}/2$. The first lattice sweep generates effective
couplings $\tilde{J}_{i}$ having 8 different values, \[
\tilde{J}_{a}=\gamma_{2}\gamma_{3}\gamma_{4}\frac{J_{a}^{5}}{J_{b}^{4}},\quad\tilde{J}_{b}=\gamma_{3}\gamma_{5}J_{a},\]
\[
\tilde{J}_{c}=\gamma_{2}\gamma_{3}\frac{J_{a}^{2}}{J_{b}},\quad\tilde{J}_{d}=\gamma_{3}\gamma_{4}\gamma_{5}\frac{J_{a}^{2}}{J_{b}},\quad\tilde{J}_{e}=\gamma_{2}^{2}\gamma_{3}\gamma_{5}\frac{J_{a}^{2}}{J_{b}},\]
\[
\tilde{J}_{f}=\gamma_{2}^{2}\gamma_{3}\gamma_{4}\gamma_{5}\frac{J_{a}^{3}}{J_{b}^{2}},\quad\tilde{J}_{g}=\gamma_{2}^{2}\gamma_{3}\gamma_{4}\frac{J_{a}^{4}}{J_{b}^{3}},\quad\tilde{J}_{h}=\gamma_{2}^{3}\gamma_{3}\gamma_{4}^{2}\frac{J_{a}^{5}}{J_{b}^{4}},\]
and whose bond distribution remains unchanged upon renormalization,
leading to the recursion relations\[
\tilde{J}_{a}^{\prime}=\gamma_{2}^{3}\frac{\tilde{J}_{a}^{2}\tilde{J}_{d}\tilde{J}_{g}}{\tilde{J}_{b}\tilde{J}_{c}^{2}},\quad\tilde{J}_{b}^{\prime}=\gamma_{3}\tilde{J}_{f},\]
\[
\tilde{J}_{c}^{\prime}=\gamma_{3}\tilde{J}_{g},\quad\tilde{J}_{d}^{\prime}=\gamma_{2}\gamma_{3}^{2}\frac{\tilde{J}_{d}\tilde{J}_{f}}{\tilde{J}_{b}},\quad\tilde{J}_{e}^{\prime}=\gamma_{2}\gamma_{3}^{2}\frac{\tilde{J}_{e}\tilde{J}_{g}}{\tilde{J}_{c}},\]
\[
\tilde{J}_{f}^{\prime}=\gamma_{2}^{2}\gamma_{3}\frac{\tilde{J}_{a}\tilde{J}_{d}\tilde{J}_{f}}{\tilde{J}_{b}\tilde{J}_{c}^{2}},\quad\tilde{J}_{g}^{\prime}=\gamma_{2}^{2}\gamma_{3}\frac{\tilde{J}_{a}\tilde{J}_{d}\tilde{J}_{g}}{\tilde{J}_{b}\tilde{J}_{c}},\quad\tilde{J}_{h}^{\prime}=\gamma_{2}^{3}\frac{\tilde{J}_{e}\tilde{J}_{f}\tilde{J}_{h}^{2}}{\tilde{J}_{b}^{2}\tilde{J}_{c}}.\]
Defining a new effective coupling $\tilde{J}_{0}=\tilde{J}_{a}\tilde{J}_{d}/\tilde{J}_{c}$,
we obtain a system of three recursion relations,\[
\tilde{J}_{0}^{\prime}=\gamma_{2}^{4}\gamma_{3}\frac{\tilde{J}_{0}^{2}\tilde{J}_{f}}{\tilde{J}_{b}^{2}},\quad\tilde{J}_{b}^{\prime}=\gamma_{3}\tilde{J}_{f},\quad\textrm{and}\quad\tilde{J}_{f}^{\prime}=\gamma_{2}^{2}\gamma_{3}\frac{\tilde{J}_{0}\tilde{J}_{f}}{\tilde{J}_{b}}.\]
With coupling ratios $\tilde{\rho}=\tilde{J}_{0}/\tilde{J}_{b}$ and
$\tilde{\sigma}=\tilde{J}_{f}/\tilde{J}_{b}$, and a gap proportional
to $\tilde{J}_{b}$, we have\[
\tilde{\rho}_{j}=\gamma_{2}^{4}\tilde{\rho}_{j-1}^{2},\quad\tilde{\sigma}_{j}=\gamma_{2}^{2}\tilde{\rho}_{j-1},\quad\textrm{and }\quad\frac{\Lambda_{j}}{\Lambda_{j-1}}=\gamma_{3}\tilde{\sigma}_{j},\]
which after eliminating $\tilde{\sigma}_{j}$ yields\[
\tilde{\rho}_{j}=\gamma_{2}^{4}\tilde{\rho}_{j-1}^{2}\quad\textrm{and}\quad\frac{\Lambda_{j}}{\Lambda_{j-1}}=\gamma_{3}\tilde{\rho}_{j}^{\nicefrac{1}{2}}.\]
Solving the recursion relations we obtain\[
\Lambda_{j}\sim r_{j}^{-\zeta}\exp\left[-\mu\left(\frac{r}{r_{0}}\right)^{\omega}\right],\]
with $\omega=\nicefrac{1}{2}$,\[
\zeta=\frac{\ln\left(\gamma_{3}/\gamma_{2}^{2}\right)}{\ln\tau_{\mathrm{rs}}}\quad\textrm{and}\quad\mu=-\frac{1}{2}\ln\left(\gamma_{2}^{2}\gamma_{4}\rho^{2}\right).\]
 So we obtain, for the whole regime $0\leq\Delta\leq1$, the dynamical
scaling form predicted for the \emph{XX} chain, reproducing the result
for the random-singlet phase. 

\begin{figure}
\includegraphics[%
  width=1.0\columnwidth]{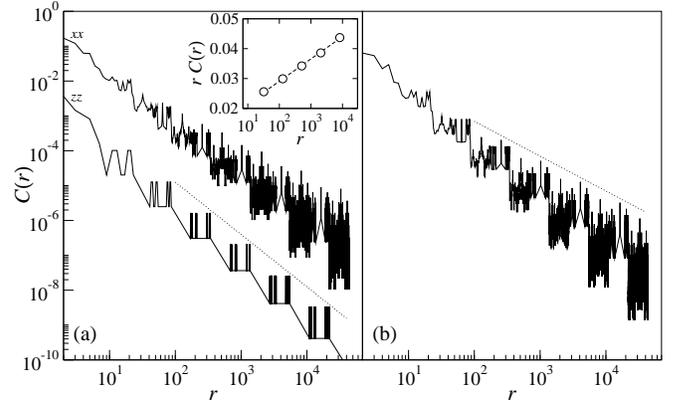}

\caption{\label{figrscorr}Ground-state correlations for chains with Rudin-Shapiro
couplings, obtained from extrapolation of numerical MDH results for
chains with $2^{16}$ to $2^{20}$ sites. (a) $C^{xx}(r)$ (upper
solid curve) and $C^{zz}(r)$ (lower solid curve) for the \emph{XX}
chain. The dotted curve is proportional to $r^{-3/2}$. (Curves offset
for clarity.) Inset: dominant $C^{xx}$ correlations, corresponding
to distances $r_{j}=2\cdot4^{j-1}$, fitted by a law of the form $rC^{xx}(r)=y_{0}+y_{1}\ln r$
(dashed curve). (b) $C(r)$ for the Heisenberg chain (solid curve).
The dotted curve is proportional to $1/r$.}
\end{figure}
For chains with RS couplings, effective-spin formation determines
the dominant ground-state correlations, but the corresponding hierarchy
is slightly different from the ones seen in Secs. \ref{sec:bm} and
\ref{sub:est}, now involving both three-spin (and some five-spin)
blocks and unrenormalized spins. As illustrated in Figs. \ref{fig:rshierab}
and \ref{fig:rshierba}, for each block renormalized in the $j$th
generation the correlation between its end spins connects a number
of order $2^{j}$ original spin pairs separated by the same distance
$r_{j}$ (the size of the block), yielding a contribution to the average
correlation in the Heisenberg chain and $C^{xx}(r_{j})$ in the \emph{XX}
chain given by\[
g_{j}=\left[\left(2c_{1,3}^{2}\right)^{j-1}+c_{2,3}^{2}\sum_{k=1}^{j-1}\left(2c_{1,3}^{2}\right)^{k-1}\right]\left|c_{0}\right|,\]
where $c_{0}$ is the correlation between end spins in a three-spin
block. For the Heisenberg chain $2c_{1,3}^{2}=\nicefrac{8}{9}<1$,
and thus \begin{equation}
C(r_{j})\sim g_{j}n_{j}\sim\frac{1}{r_{j}},\label{eq:corrheisrs}\end{equation}
 where $n_{j}\sim1/r_{j}$ is the fraction of three-spin blocks in
the $j$th generation. For the \emph{XX} chain $2\left(c_{1,3}^{xx}\right)^{2}=1$,
so that $g_{j}$ has a term proportional to $j$, and $C^{xx}(r_{j})$
carries a logarithmic correction, \begin{equation}
C^{xx}(r_{j})\sim g_{j}n_{j}\sim\left(y_{0}+y_{1}\ln r_{j}\right)r_{j}^{-1},\label{eq:corrxxrs}\end{equation}
 where $y_{0}$ and $y_{1}$ are constants. The $zz$ correlation
between end spins in a three-spin block is zero, so that the dominant
correlations correspond to spin pairs (connected through one of the
effective end spins and the middle spin) at distances $r_{j}^{\prime}=4^{j-1}\pm4^{j-2}\pm4^{j-3}\pm\cdots\pm1$,
with average $\langle r_{j}^{\prime}\rangle=4^{j-1}$, whose contribution
is given by $g_{j}^{\prime}\sim1/2^{j-1}$, since $c_{1,3}^{zz}=\nicefrac{1}{2}$.
We then have \begin{equation}
C^{zz}(r_{j}^{\prime})\sim g_{j}^{\prime}n_{j}\sim\left\langle r_{j}^{\prime}\right\rangle ^{-3/2}.\label{eq:corrzzrs}\end{equation}
 Eqs. (\ref{eq:corrxxrs}) and (\ref{eq:corrzzrs}) should be contrasted
with the random-singlet isotropic result $C(r)\sim r^{-2}$, indicating
a clear distinction between the ground-state phases induced by disorder
and aperiodicity, even in the presence of similar geometric fluctuations.
This is related to the inflation symmetry of the aperiodic sequences,
which is absent in the random-bond case (or in aperiodic systems with
random perturbations \cite{arlego02}). Its effects are exemplified
by the fractal structure of the ground-state correlations visible
in Fig. \ref{figrscorr}, which displays results from numerical implementations
of the MDH method for both \emph{XX} and Heisenberg chains, showing
conformance to the scaling forms in Eqs. (\ref{eq:corrheisrs})-(\ref{eq:corrzzrs}).
Contrary to the marginal sequences, these scaling forms should be
observed in the large-distance behavior of Rudin-Shapiro \emph{XXZ}
chains for any coupling ratio $\rho\neq1$; we expect a crossover
from the uniform to aperiodic scaling behavior as larger distances
are probed for $\rho$ close to unity. Free-fermion calculations in
the \emph{XX} limit support this picture.

\subsection{The 6-3 sequence\label{sub:63}}

\begin{figure}
\includegraphics[%
  width=1.0\columnwidth]{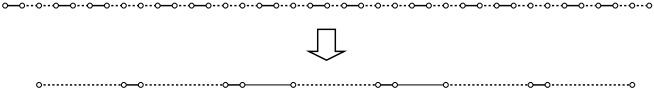}

\caption{\label{fig:63fullab}First two generations of the 6-3 sequence, discussed
in Sec. \ref{sub:63}, for $J_{a}<J_{b}$. In the second generation,
dashed lines indicate effective $\tilde{J}_{a}$ bonds, while thick
and thin solid lines denote effective $\tilde{J}_{b}$ and $\tilde{J}_{c}$
bonds. In subsequent generations, the $\tilde{J}_{c}\tilde{J}_{a}$
pairs change to $\tilde{J}_{a}\tilde{J}_{c}$, and the first $\tilde{J}_{a}$
becomes $\tilde{J}_{c}$, but the bond distribution is otherwise unchanged.}
\end{figure}
This sequence is generated by the substitution $a\rightarrow babaaa$,
$b\rightarrow baa$, and its \emph{XX} wandering exponent is $\omega=\ln2/\ln5$,
with a rescaling factor $\tau_{63}=5$. Application of the MDH scheme
leads to a fixed-point bond distribution with singlet renormalization
only, so that no effective-spin hierarchy is present. For $J_{a}<J_{b}$,
as depicted in Fig. \ref{fig:63fullab}, three effective couplings
are produced after the first lattice sweep, \[
\tilde{J}_{a}=\gamma_{2}^{2}\frac{J_{a}^{3}}{J_{b}^{2}},\quad\tilde{J}_{b}=J_{a},\quad\tilde{J}_{c}=\gamma_{2}\frac{J_{a}^{2}}{J_{b}},\]
and upon further lattice sweeps we obtain the recursion relations\[
\tilde{J}_{a}^{\prime}=\gamma_{2}^{3}\frac{\tilde{J}_{a}^{3}\tilde{J}_{c}}{\tilde{J}_{b}^{3}},\quad\tilde{J}_{b}^{\prime}=\gamma_{2}\frac{\tilde{J}_{a}\tilde{J}_{c}}{\tilde{J}_{b}},\quad\tilde{J}_{c}^{\prime}=\gamma_{2}^{2}\frac{\tilde{J}_{a}^{2}\tilde{J}_{c}}{\tilde{J}_{b}^{2}}.\]
Thus, defining the effective coupling ratios \[
\tilde{\rho}\equiv\frac{\tilde{J_{a}}}{\tilde{J}_{b}}=\gamma_{2}^{2}\left(\frac{J_{a}}{J_{b}}\right)^{2}\quad\textrm{and}\quad\tilde{\sigma}=\frac{\tilde{J}_{c}}{\tilde{J}_{b}}=\gamma_{2}\frac{J_{a}}{J_{b}},\]
 we can rewrite the recursion relations as\[
\tilde{\rho}^{\prime}=\gamma_{2}^{2}\tilde{\rho}^{2},\quad\tilde{\sigma}^{\prime}=\gamma_{2}\tilde{\rho},\quad\textrm{and}\quad\frac{\Lambda^{\prime}}{\Lambda}=\frac{\tilde{J}_{b}^{\prime}}{\tilde{J}_{b}}=\gamma_{2}\tilde{\sigma}\tilde{\rho}.\]
In the $j$th generation, we have $\tilde{\sigma}_{j}^{2}=\tilde{\rho}_{j}$,
and thus\[
\tilde{\rho}_{j+1}=\gamma_{2}^{2}\tilde{\rho}_{j}^{2}\quad\textrm{and}\quad\Lambda_{j+1}=\gamma_{2}\tilde{\rho}_{j}^{\nicefrac{3}{2}}\Lambda_{j}.\]
The length of the singlets correspond to $r_{j}=1$, $9$, $45$,
$225$,$\ldots$, so that asymptotically $r_{j}\sim r_{0}\tau^{j}$,
with $r_{0}=\nicefrac{9}{25}$ and $\tau=\tau_{63}=5$. Solving the
above recursion relations we obtain the dynamical scaling behavior,\begin{equation}
\Lambda_{j}\sim r_{j}^{-\zeta}\exp\left[-\mu\left(\frac{r}{r_{0}}\right)^{\omega}\right],\label{eq:63gsf}\end{equation}
with a wandering exponent $\omega=\ln2/\ln5\simeq0.431$ and \[
\zeta=\frac{3\ln\gamma_{2}}{\ln5}\quad\textrm{and}\quad\mu=-\frac{3}{2}\ln\left(\gamma_{2}^{4}\rho^{3}\right),\]
where $\rho=J_{a}/J_{b}$ is the original coupling ratio.

If we choose $J_{a}>J_{b}$, blocks with $2$, $3$ and $4$ spins
coupled by strong bonds appear along the chain. Effective spins are
produced by the first lattice sweep, yielding effective couplings
\[
\tilde{J}_{a}=\gamma_{2}^{3}\gamma_{3}^{2}\gamma_{4}^{3}\frac{J_{b}^{7}}{J_{a}^{6}},\quad\tilde{J}_{b}=\gamma_{2}\gamma_{3}^{2}\gamma_{4}\frac{J_{b}^{3}}{J_{a}^{2}},\quad\textrm{and}\quad\tilde{J}_{c}=\gamma_{2}^{2}\gamma_{3}^{2}\gamma_{4}^{2}\frac{J_{b}^{5}}{J_{a}^{4}},\]
whose distribution is the same as that of the third-generation bonds
for $J_{a}<J_{b}$, and which remains unchanged upon renormalization.
Thus, the scaling behavior is the same as above, but now with a `bare'
coupling ratio \[
\tilde{\rho}\equiv\frac{\tilde{J}_{a}}{\tilde{J}_{b}}=\gamma_{2}^{2}\gamma_{4}^{2}\left(\frac{J_{b}}{J_{a}}\right)^{4}.\]

\begin{figure}
\includegraphics[%
  width=1.0\columnwidth]{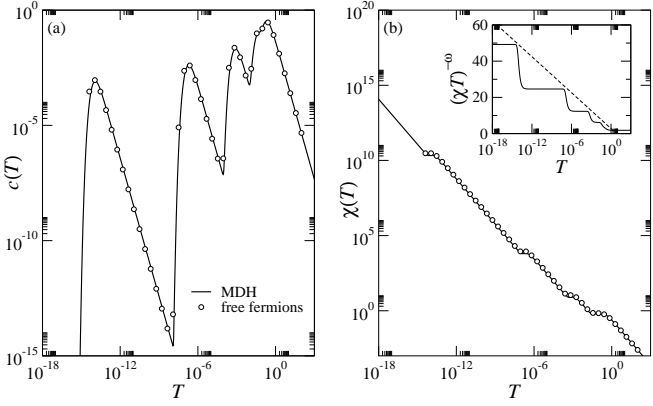}

\caption{\label{fig:63xh}Thermal dependence of the specific heat (a) and
susceptibility (b) of the \emph{XX} chain with couplings following
the 6-3 sequence of Sec. \ref{sub:63}, obtained for $J_{a}/J_{b}=\nicefrac{1}{4}$
from both numerical diagonalization of chains with $46875$ sites
(circles), and the MDH scheme (solid curves). The inset in (b) presents
a log-linear plot of $T$ versus $(\chi T)^{-\omega}$, with $\omega=\ln2/\ln5$,
showing that for $T\simeq\Lambda_{j}$, corresponding to the specific-heat
maxima, the susceptibility satisfies the scaling form $\chi\sim T^{-1}\left|\ln T\right|^{-1/\omega}$
(dashed line); at intermediate temperatures, a Curie-like behavior
is observed.}
\end{figure}
Thermodynamic properties can be estimated as in the Fibonacci case,
by using the `independent-singlet' approximation. Plots of the specific
heat $c(T$) and susceptibility $\chi(T)$ as functions of temperature
are shown in Fig. \ref{fig:63xh}, and compare quite well with results
from numerical diagonalization, even for relatively large coupling
ratios ($J_{a}/J_{b}=\nicefrac{1}{4}$). This is not surprising, given
the fact that the effective coupling ratio rapidly decreases as the
RG proceeds, even for the \emph{XX} chain. As seen in the inset of
Fig. \ref{fig:63xh}(b), at temperatures of the order of the gaps
$\Lambda_{j}$ the susceptibility follows the scaling form \[
\chi(T)\sim\frac{1}{T\left|\ln T\right|^{1/\omega}},\]
which can be readily obtained from Eq. (\ref{eq:63gsf}) by assuming
that singlet pairs are magnetically frozen, while active spins contribute
Curie terms to $\chi$. Estimates of $c(T)$ and $\chi(T)$ for chains
with anisotropies $0<\Delta\leq1$ are qualitatively identical to
the ones for the \emph{XX} chain.

\begin{figure}
\includegraphics[%
  width=1.0\columnwidth]{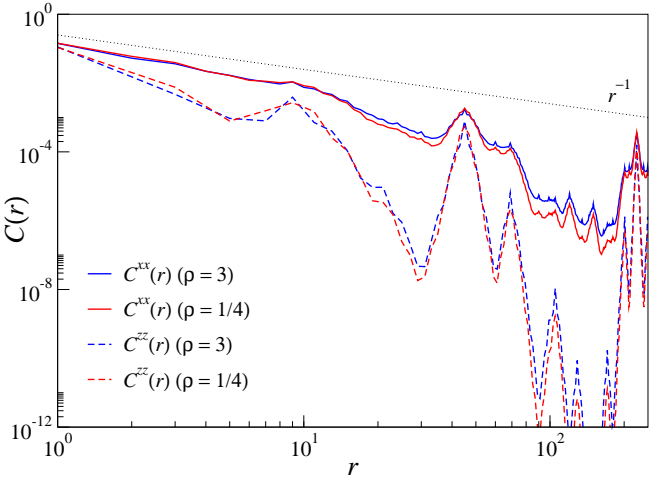}

\caption{\label{fig:s63corr}(Color online) Ground-state correlations of the
\emph{XX} chain with couplings following the 6-3 sequence of Sec.
\ref{sub:63}, for two different values of the coupling ratio $\rho=J_{a}/J_{b}$,
as obtained by numerical diagonalization of chains with $1874$ sites.
Peaks in the curves correspond to the characteristic distances $r_{j}=9$,
$45$ and $225$.}
\end{figure}
Since no effective-spin hierarchy is present, and aperiodicity is
relevant, dominant ground-state correlations, for any coupling ratio
$\rho\neq1$ and sufficiently large characteristic distances $r_{j}$,
should decay as \[
C^{xx}(r_{j})\sim C^{zz}(r_{j})\sim\frac{1}{r_{j}},\]
for all anisotropies in the regime $0\leq\Delta\leq1$. This is confirmed
in the \emph{XX} limit by numerical diagonalization, as shown in Fig.
\ref{fig:s63corr}.

\subsection{The fivefold-symmetry sequence\label{sub:ff}}

\begin{figure}
\includegraphics[%
  width=1.0\columnwidth]{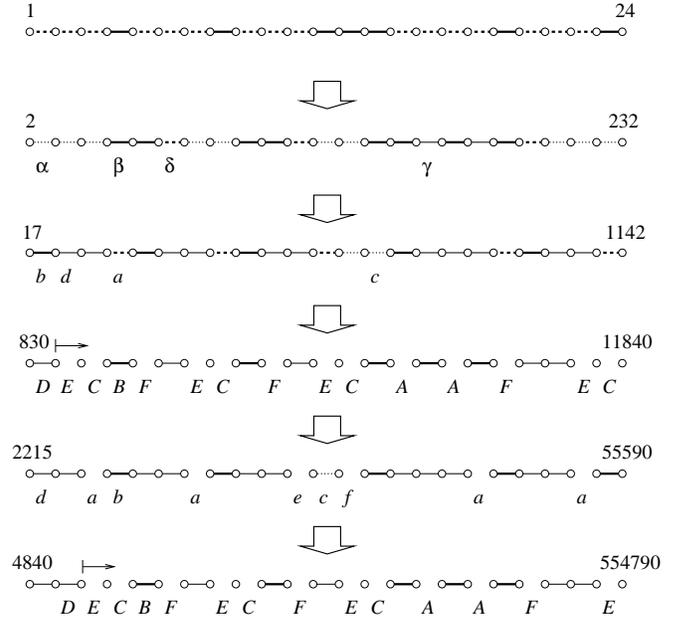}

\caption{\label{fig:fffullab}Leftmost 24 sites in the first six generations
of \emph{XXZ} chains with couplings following the fivefold-symmetry
sequence, discussed in Sec. \ref{sub:ff}, for $J_{a}<J_{b}$. The
attractor of the bond distribution is a two-cycle, reached after three
lattice sweeps. Second-generation bonds are denoted by $J_{\alpha}$
through $J_{\delta}$; third- and fifth-generation couplings are labeled
$a$ through $f$, while $A$ through $G$ label fourth- and sixth-generation
bonds. Couplings $J_{G}$ only occur much farther along the chains.
Starting from the fourth generation, lines indicate blocks to be renormalized.}
\end{figure}
The sequence produced by the substitution rule $a\rightarrow aaab$,
$b\rightarrow bba$ is related to binary tilings of the plane with
fivefold symmetry.\cite{godreche92} The (quite large) \emph{XX} rescaling
factor is $\tau_{\mathrm{ff}}=25+10\sqrt{5}\simeq47.36$, with a wandering
exponent $\omega=\ln3/\ln\tau_{\mathrm{ff}}\simeq0.285$. 

Under a numerical implementation of the MDH scheme with $J_{a}<J_{b}$,
we obtain a quite intricate pattern: after a four-bond transient produced
by the first two lattice sweeps, a two-cycle periodic attractor is
reached, where six- and seven-bond distributions alternate, as depicted
in Fig. \ref{fig:fffullab}. (With $J_{a}>J_{b},$ the same two-cycle
is reached after the first lattice sweep.) The distance between spins
connected by the strongest bonds in each generation correspond to
$r_{j}=1$, $3$, $33$, $190$, $1385$, $9050$, $\ldots$, which
asymptotically gives $r_{j+2}/r_{j}\simeq\tau_{\mathrm{ff}}$ . The
equations relating the effective couplings of the fourth and fifth
generations are\[
\tilde{J}_{a}=\gamma_{2}^{3}\gamma_{3}\frac{\tilde{J}_{A}^{2}\tilde{J}_{C}\tilde{J}_{F}}{\tilde{J}_{B}^{3}},\quad\tilde{J}_{b}=\gamma_{3}\tilde{J}_{E},\quad\tilde{J}_{c}=\gamma_{2}^{2}\frac{\tilde{J}_{A}\tilde{J}_{C}\tilde{J}_{G}}{\tilde{J}_{B}^{2}},\]
\[
\tilde{J}_{d}=\gamma_{2}^{2}\frac{\tilde{J}_{C}\tilde{J}_{E}\tilde{J}_{F}}{\tilde{J}_{B}\tilde{J}_{D}},\quad\tilde{J}_{e}=\gamma_{3}^{2}\frac{\tilde{J}_{A}^{2}\tilde{J}_{C}\tilde{J}_{G}}{\tilde{J}_{B}^{3}},\quad\tilde{J}_{f}=\gamma_{2}^{2}\gamma_{3}\frac{\tilde{J}_{A}\tilde{J}_{C}\tilde{J}_{F}}{\tilde{J}_{B}^{2}},\]
while between the couplings of the fifth and fourth generations we
have\[
\tilde{J}_{A}^{\prime}=\gamma_{2}^{2}\gamma_{3}\frac{\tilde{J}_{e}^{2}\tilde{J}_{f}}{\tilde{J}_{b}\tilde{J}_{c}},\quad\tilde{J}_{B}^{\prime}=\gamma_{2}\gamma_{3}\frac{\tilde{J}_{d}\tilde{J}_{f}}{\tilde{J}_{b}},\quad\tilde{J}_{C}^{\prime}=\gamma_{2}\frac{\tilde{J}_{e}\tilde{J}_{f}}{\tilde{J}_{b}},\]

\[
\tilde{J}_{D}^{\prime}=\gamma_{2}^{6}\gamma_{3}^{2}\frac{\tilde{J}_{a}^{2}\tilde{J}_{d}\tilde{J}_{e}\tilde{J}_{f}}{\tilde{J}_{b}^{3}\tilde{J}_{c}},\quad\tilde{J}_{E}^{\prime}=\gamma_{2}^{6}\gamma_{3}\frac{\tilde{J}_{a}^{2}\tilde{J}_{e}^{2}\tilde{J}_{f}}{\tilde{J}_{b}^{3}\tilde{J}_{c}},\]
\[
\tilde{J}_{F}^{\prime}=\gamma_{2}^{8}\gamma_{3}^{2}\frac{\tilde{J}_{a}^{3}\tilde{J}_{d}\tilde{J}_{e}\tilde{J}_{f}}{\tilde{J}_{b}^{4}\tilde{J}_{c}},\quad\tilde{J}_{G}^{\prime}=\gamma_{2}^{8}\gamma_{3}\frac{\tilde{J}_{a}^{3}\tilde{J}_{e}^{2}\tilde{J}_{f}}{\tilde{J}_{b}^{4}\tilde{J}_{c}}.\]

Eliminating the fifth-generation couplings and defining the ratios\[
\tilde{\rho}=\frac{\tilde{J}_{A}}{\tilde{J}_{B}},\quad\sigma_{1}=\frac{\tilde{J}_{C}}{\tilde{J}_{B}},\quad\sigma_{2}=\frac{\tilde{J}_{F}^{2}}{\tilde{J}_{A}\tilde{J}_{B}},\textrm{\quad and}\quad\sigma_{3}=\frac{\tilde{J}_{F}}{\tilde{J}_{E}},\]
we can write a set of four recursion relations,\[
\tilde{\rho}^{\prime}=\gamma_{2}^{3}\tilde{\rho}_{j}^{3},\quad\sigma_{1}^{\prime}=\frac{\gamma_{2}}{\gamma_{3}}\tilde{\rho}^{2},\]
\[
\sigma_{2}^{\prime}=\gamma_{2}^{22}\gamma_{3}\tilde{\rho}^{9}\left(\sigma_{1}\sigma_{3}\right)^{4},\quad\sigma_{3}^{\prime}=\gamma_{2}^{4}\gamma_{3}\sigma_{1}\sigma_{3}.\]
The gaps in successive even generations obey\[
\frac{\Lambda^{\prime}}{\Lambda}=\frac{\tilde{J}_{B}^{\prime}}{\tilde{J}_{B}}=\gamma_{2}^{5}\gamma_{3}\rho\sigma_{1}^{2}\sigma_{2},\]
and by expressing $\sigma_{1}$, $\sigma_{2}$ and $\sigma_{3}$ in
terms of $\tilde{\rho}$ we get\[
\frac{\Lambda_{j+1}}{\Lambda_{j}}=a\gamma_{2}^{20j}\tilde{\rho}_{j}^{\nicefrac{16}{3}},\]
where now $2j+2$ labels the lattice generation and $a$ is a constant
depending on the values of the coupling ratios in the fourth generation.
Solving this last equation gives\[
\Lambda_{j}\sim A^{j}B^{j^{2}}C^{3^{j}},\]
with \[
A=a\gamma_{2}^{-8},\quad B=\gamma_{2}^{20},\quad\textrm{and}\quad C=\gamma_{2}^{4/9}\tilde{\rho}_{1}^{8/9}.\]
For large enough $j$, since $j=\ln(r_{j}/r_{0})/\ln\tau_{\mathrm{{ff}}}$,
we have\[
\Lambda_{j}\sim\exp\left[-\mu\left(\frac{r_{j}}{r_{0}}\right)^{\omega}\right],\]
with\[
\mu=-\ln C\quad\textrm{and}\quad\omega=\frac{\ln3}{\ln\tau_{\mathrm{{ff}}}},\]
again obtaining, for the whole anisotropy regime $0\leq\Delta\leq1$,
the same scaling form predicted for the \emph{XX} chain.

The effective-spin hierarchy produced by the RG process is analogous
to that in the bronze-mean chains, so that ground-state correlations
behave as in Eq. (\ref{eq:bmcorr}), with $\eta=1$ in the Heisenberg
chain, but $\eta^{xx}\simeq0.884$ and $\eta^{zz}\simeq1.359$ in
the \emph{XX} limit. However, these figures are not so well reproduced
in the numerical calculations, even for chains with $N\simeq1.6\times10^{6}$
sites, most probably due to the extremely large rescaling factor.

\section{Discussion and conclusions\label{sec:disc}}

For all aperiodic sequences discussed in the previous sections, the
recursion relations for the main coupling ratio and the energy gaps
have the forms \begin{equation}
\rho_{j+1}=c\rho_{j}^{k}\quad\text{and}\quad\Lambda_{j+1}=f_{1}f_{2}^{j}\rho_{j}^{\ell}\Lambda_{j},\label{eq:rrrel}\end{equation}
 where $c$, $f_{1}$ and $f_{2}$ are $\Delta$-dependent nonuniversal
constants, and $\ell$ (a rational number) and $k$ (an integer) relate
to the number of singlets involved in determining the effective couplings.
In particular, $k$ is ultimately the difference in the number of
singlets producing the effective couplings whose ratio is $\rho<1$. 

If $k\ge2$, the recursion relation for $\rho$ always has a stable
fixed point at $\rho^{*}=0$, so that the effective coupling ratios
become exponentially small as the renormalization proceeds, indicating
that asymptotic results obtained from the MDH method should be essentially
exact. Taking into account the scaling behavior of the characteristic
distances $r_{j}\sim r_{0}\tau^{j}$, Eqs. (\ref{eq:rrrel}) lead
to the dynamical scaling form\begin{equation}
\Lambda_{j}\sim r_{j}^{-\zeta}e^{-\mu^{\prime}\ln^{2}(r_{j}/r_{0})}e^{-\mu(r_{j}/r_{0})^{\omega}}\sim e^{-\mu(r_{j}/r_{0})^{\omega}},\label{eq:reldynsc}\end{equation}
with $\zeta$ and $\mu^{\prime}$ nonuniversal constants,\[
\mu=-\frac{\ln\left(c^{\frac{\ell}{k-1}}\rho^{\ell}\right)}{k(k-1)},\]
$\rho$ being the `bare' coupling ratio, and \[
\omega=\frac{\ln k}{\ln\tau}.\]
Note that $\omega$ has the same form as the exact wandering exponent
for \emph{XX} chains with nondimerizing aperiodic couplings, given
in Eq. (\ref{eq:omegaxx}). Moreover, $\omega$ depends only on the
topology and the self-similar properties of the sequence, being independent
of the anisotropy in the regime $0\leq\Delta\leq1$. 

If $k=1$, the recursion relation for $\rho$ has a line of fixed
points, provided that $c=1$, which is generically the case in the
\emph{XX} limit; otherwise $\rho^{*}=0$ is a stable fixed point.
The general solution to Eqs. (\ref{eq:rrrel}) is\begin{equation}
\Lambda_{j}\sim r_{j}^{-\zeta(\rho)}e^{-\mu\ln^{2}(r_{j}/r_{0})},\label{eq:dynscaling}\end{equation}
where \[
\zeta(\rho)=-\frac{\ln\left(f_{1}f_{2}^{\nicefrac{1}{2}}c^{-\nicefrac{\ell}{2}}\rho^{\ell}\right)}{\ln\tau}\quad\textrm{and}\quad\mu=-\frac{\ln\left(f_{2}^{\nicefrac{1}{2}}c^{\nicefrac{\ell}{2}}\right)}{\ln^{2}\tau}.\]
Unless $f_{2}\neq1$, which, among the sequences studied here, happens
only for the relevant fivefold-symmetry sequence of Sec. \ref{sub:ff},
$\mu$ is zero in the \emph{XX} limit. This means that we can identify
a nonuniversal dynamical exponent $z=\zeta(\rho)$, and the scaling
behavior of thermodynamic properties depends on the coupling ratio
for the whole anisotropy regime $0\leq\Delta<1$. In the Heisenberg
limit ($\Delta=1$), unless $\mu=0$, as in the marginal tripling
sequence of Sec. \ref{sub:smt}, Eq. (\ref{eq:dynscaling}) describes
a weakly exponential dynamic scaling. In this case, aperiodicity can
be viewed as a marginally relevant operator ($\omega\rightarrow0^{+}$)
in the renormalization-group sense.

These results strongly suggest that low-temperature thermodynamic
properties of any antiferromagnetic \emph{XXZ} chain with anisotropies
intermediate between the \emph{XX} and Heisenberg limits, and couplings
following a given binary aperiodic sequence, can be classified according
to a single wandering exponent $\omega$, which is known exactly for
\emph{XX} chains. This generalizes what happens in random-bond \emph{XXZ}
chains (for which $\omega=\nicefrac{1}{2}$), where thermodynamic
properties in the anisotropy regime $-\nicefrac{1}{2}\leq\Delta\leq1$
are those characterizing the random-singlet phase.\cite{doty92,fisher94}
Note that, although the above classification seems to imply an anisotropy-independent
critical value $\omega_{c}=0$ for the relevance of aperiodic fluctuations
on the low-temperature behavior of \emph{XXZ} chains, it does not
show that $\omega$ plays the role of a genuine wandering exponent,
in the sense that fluctuations scale as $g\sim N^{\omega}$, for general
easy-plane anisotropies. In any case, due to the fact that the critical
exponents (including the correlation-length exponent $\nu$) of the
uniform \emph{XXZ} chain are known to vary with the anisotropy along
the whole critical line $-1\leq\Delta\leq1$,\cite{baxter72,luther75}
it remains an open question how the present results fit into the framework
of the Harris-Luck criterion. 

Of course, Eqs. (\ref{eq:rrrel}) are valid for all anisotropies $0\leq\Delta\leq1$
only if the bond distribution generated by the MDH method is independent
of $\Delta$. This is certainly the case for strong enough modulation.
(How strong this modulation has to be depends on the various block
sizes produced by the sequence.) However, from numerical implementations
of the method, we find that, even when the blocks selected for renormalization
in the first few lattice sweeps depend on $\Delta$, a universal distribution
is eventually reached, in much the same way as when we choose $J_{a}>J_{b}$
instead of $J_{a}<J_{b}$. Thus, we expect that, for general binary
substitution rules inducing relevant aperiodicity, the scaling form
in Eq. (\ref{eq:reldynsc}) holds for all coupling ratios $\rho\neq1$.

An approximate picture of the ground state and of the lowest excitations
in the presence of aperiodic couplings can also be deduced from the
MDH scheme, and is revealed by the behavior of the pair correlation
functions. As the energy scale is reduced, two types of behavior can
be identified: either the RG process produces a hierarchy of singlets
(as in the Fibonacci, silver-mean, marginal-tripling, and 6-3 sequences),
or a hierarchy of effective spins (as in the bronze-mean, spin-triple,
Rudin-Shapiro, and fivefold-symmetry sequences). The first type reveals
a kind of self-similar, `aperiodic-singlet' phase, from which (singlet-triplet)
excitations involve strongly coupled pairs composed of spins separated
by well-defined characteristic distances. In the second type, since
the number of spins contributing to an effective spin increases exponentially
along the hierarchy, excitations of a certain energy involve spins
separated by a wide range of distances, giving rise to a fractal structure
of the correlation functions. Notice that, contrary to the finite
temperature behavior, there is no relation between the ground-state
properties and the marginal or relevant character of the aperiodicity.

For aperiodic sequences inducing strictly marginal fluctuations, we
could account for the nonuniversality of the correlation-function
decay exponents by a numerical calculation based on a second-order
expansion of the ground-state vectors. This compares quite well with
results from numerical diagonalization in the \emph{XX} limit, which
indeed show that the zeroth-order MDH predictions are reproduced in
the strong-modulation regime.

The results on relevant aperiodic couplings show that geometrical
fluctuations, measured by the wandering exponent $\omega$, are not
determinant for ground-state properties, although they control the
low-energy scaling of thermodynamic quantities. In particular, both
random bonds and Rudin-Shapiro couplings are characterized by $\omega=\nicefrac{1}{2}$;
however, correlations in the random-singlet phase are entirely different
from those in \emph{XXZ} chains with Rudin-Shapiro couplings. This
is a consequence of the inflation symmetry induced by substitution
rules, which is clearly absent in random chains. (Analogously, comparative
studies\cite{igloi98,igloi98b} between random-bond and Rudin-Shapiro
quantum Ising chains show that, although the corresponding scaling
properties are similar at the critical point, only randomness is capable
of producing the off-critical Griffiths singularities.\cite{griffiths69,fisher92,fisher95})
Nevertheless, aperiodic and random \emph{XXZ} chains share the feature
that average and typical behavior are strikingly distinct, and that
average correlations decay as power laws. Finally, aperiodic ground-state
phases are unstable towards random perturbations, which break inflation
symmetry, and the random-singlet behavior is ultimately recovered.\cite{arlego02}

\begin{acknowledgments}
This work has been supported by the Brazilian agencies CAPES and FAPESP.
The author is indebted to T. A. S. Haddad, E. Miranda, J. A. Hoyos,
A. P. S. de Moura, F. C. Alcaraz, and S. R. Salinas for helpful conversations.
\end{acknowledgments}
\appendix

\section{Renormalization of multi-spin blocks\label{sec:multi}}

In this Appendix, we derive the expressions for the renormalized coupling
constants used in the extension of the Ma-Dasgupta-Hu method to \emph{XXZ}
chains with aperiodic couplings. 

Contrary to the random-bond chains discussed in Sec. \ref{sec:randombond},
when couplings follow aperiodic sequences generated by inflation rules
we generally need to consider spin blocks with more than one strong
bond, and thus more than two spins. For instance, in the Fibonacci
sequence with $J_{a}>J_{b}$ (see Fig. \ref{fig:fibfullba}) there
appear blocks with one or two $J_{a}$ bonds. Since we assume that
all couplings are antiferromagnetic, the local ground state is a singlet
for blocks with an even number of spins, but a doublet if the blocks
contain an odd number of spins.\cite{lieb62} 

Let us consider a block with $n$ spins $S_{1}$ through $S_{n}$
connected by equal bonds $J_{0}$, with anisotropy $\Delta_{0}$.
This is described by the local Hamiltonian\[
H_{0}=J_{0}\sum_{j=1}^{n-1}\left(\mathbf{S}_{j}\cdot\mathbf{S}_{j+1}\right)_{\Delta_{0}},\]
 where we introduced the notation\[
\left(\mathbf{S}_{i}\cdot\mathbf{S}_{j}\right)_{\Delta}\equiv S_{i}^{x}S_{j}^{x}+S_{i}^{y}S_{j}^{y}+\Delta S_{i}^{z}S_{j}^{z}.\]
The gap $\Lambda_{0}$ between the ground-state energy of the block
and its lowest excited multiplet depends on $J_{0}$ and $\Delta_{0}$.
For two-spin and three-spin blocks we have\[
\Lambda_{0}^{(2)}=\frac{1+\Delta_{0}}{2}J_{0}\quad\textrm{and}\quad\Lambda_{0}^{(3)}=\frac{1}{4}\left(\Delta_{0}+\sqrt{\Delta_{0}^{2}+8}\right)J_{0}.\]
We define the strongest bonds in the chain as those producing spin
blocks with the largest gaps $\Lambda_{0}$. 

An $n$-spin block to be renormalized is connected to its neighboring
spins $S_{l}$ and $S_{r}$ by weaker bonds $J_{l}$ and $J_{r}$.
The relevant part of the chain Hamiltonian is \[
H=H_{0}+H_{lr},\]
with\begin{equation}
H_{lr}=J_{l}\left(\mathbf{S}_{l}\cdot\mathbf{S}_{1}\right)_{\Delta_{l}}+J_{r}\left(\mathbf{S}_{n}\cdot\mathbf{S}_{r}\right)_{\Delta_{r}}.\label{eq:hlr}\end{equation}
The idea of the MDH method is to obtain recursion relations for the
couplings by treating $H_{lr}$ as a perturbation to $H_{0}$.

\begin{figure}
\includegraphics[%
  width=0.50\columnwidth]{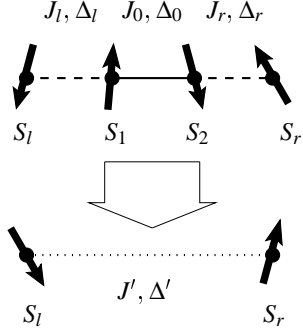}

\caption{\label{fig:appfig1}Renormalization step involving the decimation
of a two-spin block.}
\end{figure}
If $n$ is even (as in the two-spin case shown in Fig. \ref{fig:appfig1}),
the ground state of $H_{0}$ is a singlet, which we denote by $\left|\Psi_{0}\right\rangle $,
with a corresponding energy $E_{0}$. In the space of this singlet,
the states of $S_{l}$ and $S_{r}$ are arbitrary. In the space spanned
by the eigenstates $\left|\Psi_{i}\right\rangle $ of $H_{0}$ (with
energies $E_{i}$) and the states $\left|m_{l},m_{r}\right\rangle \equiv\left|m_{l}\right\rangle \otimes\left|m_{r}\right\rangle $
of $S_{l.r}$ ($m_{l,r}=\pm\nicefrac{1}{2}$), the states $\left|g(m_{l},m_{r})\right\rangle \equiv\left|m_{l},m_{r}\right\rangle \otimes\left|\Psi_{0}\right\rangle $
are degenerate. The first-order perturbative corrections to the ground-state
energy $E_{0}$ are zero, but the second-order corrections are given
by the eigenvalues of the matrix\[
V_{m_{l},m_{r};m_{l}^{\prime},m_{r}^{\prime}}=\sum_{e}\frac{\left\langle g(m_{l},m_{r})\left|H_{lr}\right|e\right\rangle \left\langle e\left|H_{lr}\right|g(m_{l}^{\prime},m_{r}^{\prime})\right\rangle }{E_{0}-E_{i}},\]
where the summation spans the excited states $\left|e\right\rangle \equiv\left|m_{l}^{\prime\prime},m_{r}^{\prime\prime}\right\rangle \otimes\left|\Psi_{i}\right\rangle $
($i=1,\ldots,2^{n}-1$). In terms of the raising and lowering operators
$S^{\pm}=S^{x}\pm iS^{y}$ we have \begin{equation}
\left(\mathbf{S}_{i}\cdot\mathbf{S}_{j}\right)_{\Delta}\equiv\frac{1}{2}\left(S_{i}^{+}S_{j}^{-}+S_{i}^{-}S_{j}^{+}\right)+\Delta S_{i}^{z}S_{j}^{z},\label{eq:sijpm}\end{equation}
and a little algebra shows that\begin{widetext}\begin{eqnarray}
V_{m_{l},m_{r};m_{l}^{\prime},m_{r}^{\prime}} & = & \frac{1}{4}J_{l}J_{r}\left\langle m_{l},m_{r}\left|S_{l}^{+}S_{r}^{-}\right|m_{l}^{\prime},m_{r}^{\prime}\right\rangle \sum_{i\neq0}\frac{\left\langle \Psi_{0}\left|S_{1}^{-}\right|\Psi_{i}\right\rangle \left\langle \Psi_{i}\left|S_{n}^{+}\right|\Psi_{0}\right\rangle +\left\langle \Psi_{0}\left|S_{n}^{+}\right|\Psi_{i}\right\rangle \left\langle \Psi_{i}\left|S_{1}^{-}\right|\Psi_{0}\right\rangle }{E_{0}-E_{i}}\nonumber \\
 & + & \frac{1}{4}J_{l}J_{r}\left\langle m_{l},m_{r}\left|S_{l}^{-}S_{r}^{+}\right|m_{l}^{\prime},m_{r}^{\prime}\right\rangle \sum_{i\neq0}\frac{\left\langle \Psi_{0}\left|S_{1}^{+}\right|\Psi_{i}\right\rangle \left\langle \Psi_{i}\left|S_{n}^{-}\right|\Psi_{0}\right\rangle +\left\langle \Psi_{0}\left|S_{n}^{-}\right|\Psi_{i}\right\rangle \left\langle \Psi_{i}\left|S_{1}^{+}\right|\Psi_{0}\right\rangle }{E_{0}-E_{i}}\nonumber \\
 & + & \Delta_{l}\Delta_{r}J_{l}J_{r}\left\langle m_{l},m_{r}\left|S_{l}^{z}S_{r}^{z}\right|m_{l}^{\prime},m_{r}^{\prime}\right\rangle \sum_{i\neq0}\frac{\left\langle \Psi_{0}\left|S_{1}^{z}\right|\Psi_{i}\right\rangle \left\langle \Psi_{i}\left|S_{n}^{z}\right|\Psi_{0}\right\rangle +\left\langle \Psi_{0}\left|S_{n}^{z}\right|\Psi_{i}\right\rangle \left\langle \Psi_{i}\left|S_{1}^{z}\right|\Psi_{0}\right\rangle }{E_{0}-E_{i}}.\label{eq:vlr}\end{eqnarray}
 \end{widetext}Since the first two terms on the right-hand side of
Eq. (\ref{eq:vlr}) are complex conjugates, and noting that $E_{0}-E_{i}$
is proportional to $J_{0}$, we can write\begin{eqnarray*}
V_{m_{l},m_{r};m_{l}^{\prime},m_{r}^{\prime}} & = & \gamma_{n}\frac{J_{l}J_{r}}{J_{0}}\left\langle m_{l},m_{r}\left|S_{l}^{x}S_{r}^{x}+S_{l}^{y}S_{r}^{y}\right|m_{l}^{\prime},m_{r}^{\prime}\right\rangle \\
 & + & \gamma_{n}\delta_{n}\Delta_{l}\Delta_{r}\frac{J_{l}J_{r}}{J_{0}}\left\langle m_{l},m_{r}\left|S_{l}^{z}S_{r}^{z}\right|m_{l}^{\prime},m_{r}^{\prime}\right\rangle ,\end{eqnarray*}
where $\gamma_{n}$ and $\delta_{n}$ depend on $\Delta_{0}$ (with
$\delta_{n}=1$ for $\Delta_{0}=1$, where $\textrm{SU}(2)$ symmetry
is recovered). The above matrix elements are precisely the ones corresponding
to the Hamiltonian\[
H^{\prime}=J^{\prime}\left(S_{l}^{x}S_{r}^{x}+S_{l}^{y}S_{r}^{y}+\Delta^{\prime}S_{l}^{z}S_{r}^{z}\right),\]
with the effective parameters $J^{\prime}$ and $\Delta^{\prime}$
given by\begin{equation}
J^{\prime}=\gamma_{n}\frac{J_{l}J_{r}}{J_{0}}\quad\textrm{and}\quad\Delta^{\prime}=\delta_{n}\Delta_{l}\Delta_{r}.\label{eq:rreven}\end{equation}
For two-spin blocks we have\[
\gamma_{2}=\frac{1}{1+\Delta_{0}}\quad\textrm{and}\quad\delta_{2}=\frac{1+\Delta_{0}}{2}.\]
For larger blocks the parameters can be evaluated numerically as a
function of $\Delta_{0}$; however, for four-spin blocks we can analytically
determine $\gamma_{4}=1$ in the \emph{XX} limit and $\gamma_{4}=\nicefrac{2}{3}-\nicefrac{\sqrt{3}}{6}\simeq0.378$
in the Heisenberg chain.

\begin{figure}
\includegraphics[%
  width=0.50\columnwidth]{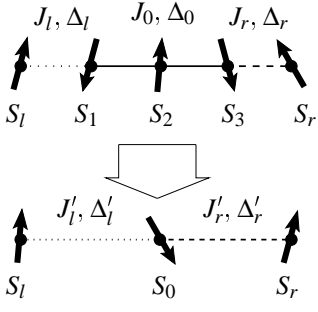}

\caption{\label{fig:appfig2}Renormalization step involving a three-spin block.}
\end{figure}
If $n$ is odd (as in the three-spin case shown in Fig. \ref{fig:appfig2}),
$H_{0}$ has two degenerate ground states, which we denote by $\left|\Psi_{0}^{\pm}\right\rangle $.
These can be associated with an effective spin-$\nicefrac{1}{2}$
$S_{0}$, whose states can be described by the azimuthal quantum number
$m_{0}$, so that $\left|m_{0}=\pm\nicefrac{1}{2}\right\rangle =\left|\Psi_{0}^{\pm}\right\rangle $.
In the space $\mathbb{H}$ spanned by the states of $S_{0}$, $S_{l}$
and $S_{r}$, the states $\left|m_{l},m_{r},m_{0}\right\rangle \equiv\left|m_{l}\right\rangle \otimes\left|m_{r}\right\rangle \otimes\left|m_{0}\right\rangle $
are degenerate. The introduction of $H_{lr}$ lifts this degeneracy,
and we expect that, to order $J_{l,r}/J_{0}$, perturbation theory
leads to an effective Hamiltonian $H^{\prime}$, with matrix elements
given (apart from a constant) by\[
H_{m_{l},m_{r},m_{0};m_{l}^{\prime},m_{r}^{\prime},m_{0}^{\prime}}^{\prime}=\left\langle m_{l},m_{r},m_{0}\left|H_{lr}\right|m_{l}^{\prime},m_{r}^{\prime},m_{0}^{\prime}\right\rangle .\]
Restricting ourselves to the space $\mathbb{H}$, it is possible to
write \[
H^{\prime}=J_{l}^{\prime}\left(\mathbf{S}_{l}\cdot\mathbf{S}_{0}\right)_{\Delta_{l}^{\prime}}+J_{r}^{\prime}\left(\mathbf{S}_{0}\cdot\mathbf{S}_{r}\right)_{\Delta_{r}^{\prime}},\]
provided \begin{equation}
J_{l}\left(\mathbf{S}_{l}\cdot\mathbf{S}_{1}\right)_{\Delta_{l}}=J_{l}^{\prime}\left(\mathbf{S}_{l}\cdot\mathbf{S}_{0}\right)_{\Delta_{l}^{\prime}}\ \ \label{eq:jll}\end{equation}
and\[
J_{r}\left(\mathbf{S}_{n}\cdot\mathbf{S}_{r}\right)_{\Delta_{r}}=J_{r}^{\prime}\left(\mathbf{S}_{0}\cdot\mathbf{S}_{r}\right)_{\Delta_{r}^{\prime}}.\]
We now expand Eq. (\ref{eq:jll}) with the help of Eq. (\ref{eq:sijpm}),
and notice that\[
\left\langle m_{l},m_{r},m_{0}\left|S_{l}^{+}S_{0}^{-}\right|m_{l}^{\prime},m_{r}^{\prime},m_{0}^{\prime}\right\rangle =\delta_{m_{l}^{\prime},m_{l}-1}\delta_{m_{r}^{\prime},m_{r}}\delta_{m_{0}^{\prime},m_{0}+1},\]
\[
\left\langle m_{l},m_{r},m_{0}\left|S_{l}^{+}S_{1}^{-}\right|m_{l}^{\prime},m_{r}^{\prime},m_{0}^{\prime}\right\rangle =\delta_{m_{l}^{\prime},m_{l}-1}\delta_{m_{r}^{\prime},m_{r}}\left\langle m_{0}\left|S_{1}^{-}\right|m_{0}^{\prime}\right\rangle ,\]
\[
\left\langle m_{l},m_{r},m_{0}\left|S_{l}^{z}S_{0}^{z}\right|m_{l}^{\prime},m_{r}^{\prime},m_{0}^{\prime}\right\rangle =m_{l}m_{0}\delta_{m_{l}^{\prime},m_{l}}\delta_{m_{r}^{\prime},m_{r}}\delta_{m_{0}^{\prime},m_{0}},\]
\[
\left\langle m_{l},m_{r},m_{0}\left|S_{l}^{z}S_{1}^{z}\right|m_{l}^{\prime},m_{r}^{\prime},m_{0}^{\prime}\right\rangle =m_{l}\delta_{m_{l}^{\prime},m_{l}}\delta_{m_{r}^{\prime},m_{r}}\left\langle m_{0}\left|S_{1}^{z}\right|m_{0}^{\prime}\right\rangle ,\]
$\delta_{i,j}$ being the Kronecker symbol. By the Wigner-Eckart theorem,
we can write\[
\left\langle m_{0}\left|S_{1}^{-}\right|m_{0}^{\prime}\right\rangle =\gamma_{n}\delta_{m_{0}^{\prime},m_{0}+1},\]
\[
\left\langle m_{0}\left|S_{1}^{z}\right|m_{0}^{\prime}\right\rangle =\left(\delta_{n}\gamma_{n}\right)m_{0}\delta_{m_{0}^{\prime},m_{0}},\]
with $\gamma_{n}$ and $\delta_{n}$ depending on $\Delta_{0}$, and
we formally obtain the renormalized parameters\begin{equation}
J_{l}^{\prime}=\gamma_{n}J_{l}\quad\textrm{and}\quad\Delta_{l}^{\prime}=\delta_{n}\Delta_{l}.\label{eq:rrodd1}\end{equation}
Analogously, by symmetry we have\begin{equation}
J_{r}^{\prime}=\gamma_{n}J_{r}\quad\textrm{and}\quad\Delta_{r}^{\prime}=\delta_{n}\Delta_{r}.\label{eq:rrodd2}\end{equation}
For three-spin blocks we obtain\[
\gamma_{3}=\frac{\left(\Delta_{0}+\sqrt{\Delta_{0}^{2}+8}\right)}{2+\frac{1}{4}\left(\Delta_{0}+\sqrt{\Delta_{0}^{2}+8}\right)^{2}}\]
and\[
\delta_{3}=\frac{1}{4}\left(\Delta_{0}+\sqrt{\Delta_{0}^{2}+8}\right),\]
while for larger blocks the parameters can be calculated numerically.
In particular, for five-spin blocks we have, in the \emph{XX} limit
(for which analytical results are available), $\gamma_{5}=\tfrac{\sqrt{3}}{3}\simeq0.577,$
and in the Heisenberg chain $\gamma_{5}\simeq0.512$.

In blocks with an odd number of spins, the original spins $S_{i}$
($i=1,\ldots,n$) are represented by the effective spin $S_{0}$,
with `weights' given by the coefficients $c_{i,n}^{x}$ and $c_{i,n}^{z}$,
defined through the operator identities (valid in $\mathbb{H}$)\[
S_{i}^{x}=c_{i,n}^{x}S_{0}^{x}\quad\textrm{and}\quad S_{i}^{z}=c_{i,n}^{z}S_{0}^{z}.\]
These are useful in the calculation of correlation functions. Note
that $c_{1,n}^{x}=c_{n,n}^{x}=\gamma_{n}$ and $c_{1,n}^{z}=c_{n,n}^{z}=\delta_{n}\gamma_{n}$.
For three-spin blocks we have\[
c_{2,3}^{x}=-\frac{1}{1+\frac{1}{8}\left(\Delta_{0}+\sqrt{\Delta_{0}^{2}+8}\right)^{2}}\]
and\[
c_{2,3}^{z}=\frac{1}{4}\Delta_{0}\left(\Delta_{0}+\sqrt{\Delta_{0}^{2}+8}\right)c_{2,3}^{x}.\]

Equations (\ref{eq:rreven}), (\ref{eq:rrodd1}) and (\ref{eq:rrodd2})
constitute the recursion relations defining the RG steps in the MDH
scheme.

\section{Second-order calculation of correlation functions\label{sec:corr2}}

Let us assume that a two-spin block, as the one shown in Fig. \ref{fig:appfig1},
is selected for renormalization at some point of the RG process. In
terms of the states of $S_{1}$ and $S_{2}$, the eigenstates of the
block Hamiltonian $H_{0}$, with the corresponding energies, are \[
\left|\Psi_{0}\right\rangle =\tfrac{1}{\sqrt{2}}\left(\left|\uparrow\downarrow\right\rangle -\left|\downarrow\uparrow\right\rangle \right),\quad E_{0}=-\left(\tfrac{1}{2}+\tfrac{1}{4}\Delta_{0}\right)J_{0},\]
\[
\left|\Psi_{1}\right\rangle =\left|\uparrow\uparrow\right\rangle ,\quad\left|\Psi_{2}\right\rangle =\left|\downarrow\downarrow\right\rangle ,\quad E_{1}=E_{2}=\tfrac{1}{4}\Delta_{0}J_{0},\]
and \[
\left|\Psi_{3}\right\rangle =\tfrac{1}{\sqrt{2}}\left(\left|\uparrow\downarrow\right\rangle +\left|\downarrow\uparrow\right\rangle \right),\quad E_{3}=\left(\tfrac{1}{2}-\tfrac{1}{4}\Delta_{0}\right)J_{0}.\]
The connection between the two-spin block and the rest of the chain,
through the neighboring spins $S_{l}$ and $S_{r}$, is described
by the Hamiltonian $H_{lr}$ in Eq. (\ref{eq:hlr}).

Denoting by $\left|A_{i}\right\rangle $ the states of all other spins
in the chain, and assuming that in the thermodynamic limit there is
a unique ground state $\left|A_{0}\right\rangle $, the ground state
of the whole chain can be written, at zeroth-order in perturbation
theory, as $\left|g_{0}\right\rangle =\left|A_{0},\Psi_{0}\right\rangle $.
Up to second order in $J_{l,r}/J_{0}$ we obtain a corrected state\begin{widetext}
\begin{eqnarray}
\left|g\right\rangle  & = & \left|g_{0}\right\rangle +\sum_{i}\sum_{k\neq0}\left|A_{i},\Psi_{k}\right\rangle \frac{\left\langle A_{i},\Psi_{k}\left|H_{lr}\right|A_{0},\Psi_{0}\right\rangle }{E_{0}-E_{k}}+\sum_{i,j}\sum_{k,\ell\neq0}\left|A_{i},\Psi_{k}\right\rangle \frac{\left\langle A_{i},\Psi_{k}\left|H_{lr}\right|A_{j},\Psi_{\ell}\right\rangle \left\langle A_{j},\Psi_{\ell}\left|H_{lr}\right|A_{0},\Psi_{0}\right\rangle }{\left(E_{0}-E_{k}\right)\left(E_{0}-E_{\ell}\right)}\nonumber \\
 & - & \sum_{i,j}\sum_{k\neq0}\left|A_{i},\Psi_{k}\right\rangle \frac{\left\langle A_{i},\Psi_{k}\left|H_{lr}\right|A_{j},\Psi_{0}\right\rangle \left\langle A_{j},\Psi_{0}\left|H_{lr}\right|A_{0},\Psi_{0}\right\rangle }{\left(E_{0}-E_{k}\right)^{2}}.\label{eq:g2}\end{eqnarray}
\end{widetext}A second-order estimate of the expectation value of
any operator $O$ is simply given by\[
\left\langle O\right\rangle _{g}\equiv\frac{\left\langle g\left|O\right|g\right\rangle }{\left\langle \left.g\right|g\right\rangle }.\]

For the calculation of correlation functions involving spins in the
block, we write $O=O_{\Psi}O_{A}$, where $O_{\Psi}$ and $O_{A}$
are operators acting on the subspaces defined by the states $\left|\Psi_{i}\right\rangle $
and $\left|A_{i}\right\rangle $, respectively. Expanding Eq. (\ref{eq:g2}),
we obtain an expression for $\left\langle g\left|O\right|g\right\rangle $
with terms containing combinations such as $\left\langle \Psi_{i}\left|O_{\Psi}\right|\Psi_{j}\right\rangle $
and $\left\langle A_{0}\left|S_{l}^{+}O_{A}\right|A_{0}\right\rangle $,
which is rather cumbersome to write here. As examples of the final
results obtained in the Heisenberg limit, we have\[
\left\langle \left.g\right|g\right\rangle \equiv g^{-1}=1+\tfrac{3}{16}\frac{J_{l}^{2}+J_{r}^{2}}{J_{0}^{2}}-\tfrac{1}{2}\frac{J_{l}J_{r}}{J_{0}^{2}}\left\langle A_{0}\left|\mathbf{S}_{l}\cdot\mathbf{S}_{r}\right|A_{0}\right\rangle ,\]
\[
\left\langle \mathbf{S}_{1}\cdot\mathbf{S}_{2}\right\rangle _{g}=-\tfrac{3}{4}g\left(1-\tfrac{1}{16}\frac{J_{l}^{2}+J_{r}^{2}}{J_{0}^{2}}+\tfrac{1}{6}\frac{J_{l}J_{r}}{J_{0}^{2}}\left\langle A_{0}\left|\mathbf{S}_{l}\cdot\mathbf{S}_{r}\right|A_{0}\right\rangle \right),\]
\begin{eqnarray*}
\left\langle \mathbf{S}_{l}\cdot\mathbf{S}_{1}\right\rangle _{g}= & \tfrac{1}{2}g & \left[\left(\tfrac{1}{3}+\tfrac{1}{4}\frac{J_{r}+2J_{l}}{J_{0}}\right)\frac{J_{r}}{J_{0}}\left\langle A_{0}\left|\mathbf{S}_{l}\cdot\mathbf{S}_{r}\right|A_{0}\right\rangle \right.\\
 &  & -\left.\tfrac{1}{4}\left(1+\tfrac{3}{4}\frac{J_{l}}{J_{0}}\right)\frac{J_{l}}{J_{0}}\right],\end{eqnarray*}
and\begin{eqnarray*}
\left\langle \mathbf{S}_{n}\cdot\mathbf{S}_{1}\right\rangle _{g}= & -\tfrac{1}{2}g & \left[\left(1+\tfrac{1}{4}\frac{J_{l}}{J_{0}}\right)\frac{J_{l}}{J_{0}}\left\langle A_{0}\left|\mathbf{S}_{n}\cdot\mathbf{S}_{l}\right|A_{0}\right\rangle \right.\\
 &  & -\left.\left(1+\tfrac{3}{4}\frac{J_{r}}{J_{0}}\right)\frac{J_{r}}{J_{0}}\left\langle A_{0}\left|\mathbf{S}_{n}\cdot\mathbf{S}_{r}\right|A_{0}\right\rangle \right],\end{eqnarray*}
$S_{n}$ being any spin other than $S_{l}$, $S_{r}$, $S_{1}$ and
$S_{2}$. These expressions depend explicitly on expectation values
like $\left\langle A_{0}\left|\mathbf{S}_{l}\cdot\mathbf{S}_{r}\right|A_{0}\right\rangle $,
which on their turn depend on expectation values involving spins neighboring
the blocks in which $S_{l}$ and $S_{r}$ will be decimated. As the
RG proceeds, this generates a hierarchical structure, which can be
solved backwards by assuming that the correlation between the spins
in the very last block to be renormalized takes it zeroth-order value.
It is interesting to notice that the correlation between two spins
which are not decimated in the same block is at most of order $J_{l,r}/J_{0}$. 

Similarly, in the \emph{XX} limit we have, for instance,\[
\left\langle \left.g\right|g\right\rangle \equiv g^{-1}=1+\tfrac{1}{2}\frac{J_{l}^{2}+J_{r}^{2}}{J_{0}^{2}}-4\frac{J_{l}J_{r}}{J_{0}^{2}}\left\langle A_{0}\left|S_{l}^{x}S_{r}^{x}\right|A_{0}\right\rangle ,\]

\[
\left\langle S_{1}^{x}S_{2}^{x}\right\rangle _{g}=-\tfrac{1}{4}g,\]
\[
\left\langle S_{1}^{z}S_{2}^{z}\right\rangle _{g}=-\tfrac{1}{4}g\left(1-\tfrac{1}{2}\frac{J_{l}^{2}+J_{r}^{2}}{J_{0}^{2}}+4\frac{J_{l}J_{r}}{J_{0}^{2}}\left\langle A_{0}\left|S_{l}^{x}S_{r}^{x}\right|A_{0}\right\rangle \right),\]
\[
\left\langle S_{n}^{x}S_{1,2}^{x}\right\rangle _{g}=-g\left(\frac{J_{l,r}}{J_{0}}\left\langle A_{0}\left|S_{n}^{x}S_{l,r}^{x}\right|A_{0}\right\rangle -\frac{J_{r,l}}{J_{0}}\left\langle A_{0}\left|S_{n}^{x}S_{r,l}^{x}\right|A_{0}\right\rangle \right),\]
\[
\left\langle S_{n}^{z}S_{1,2}^{z}\right\rangle _{g}=g\frac{J_{r,l}^{2}}{J_{0}^{2}}\left\langle A_{0}\left|S_{n}^{z}S_{r,l}^{z}\right|A_{0}\right\rangle .\]
Notice that expressions for the $zz$ correlations may involve other
expectation values of both $xx$ and $zz$ correlations.

%\bibliographystyle{apsrev}
%\bibliography{mestrado}

\end{document}